\documentclass[10pt,fleqn]{article}
\usepackage[utf8]{inputenc}
\usepackage[T1]{fontenc}
\usepackage[top=1in,bottom=1in,left=1in,right=1in]{geometry}
\usepackage{authblk}
\usepackage{amsmath,amsfonts,amssymb}
\numberwithin{equation}{section}
\newcommand{\ket}[1]{\rvert#1\rangle}

\title{Generating functions for the $\mathfrak{osp}(1|2)$\\Clebsch-Gordan coefficients}
\author[1]{Geoffroy Bergeron}
\author[1]{Luc Vinet}
\affil[1]{Centre de recherches math\'ematiques, Universit\'e de Montr\'eal, P.O. Box 6128, Centre-ville Station, Montr\'eal, Canada H3C 3J7}

\date{}
\begin{document}
\maketitle
\thispagestyle{empty}
\hrule
\begin{abstract}\noindent
Generating functions for Clebsch-Gordan coefficients of $\mathfrak{osp}(1|2)$ are derived. These coefficients are expressed as $q \rightarrow -1$ limits of the dual  $q$-Hahn polynomials. The generating functions are obtained using two different approaches respectively based on the coherent-state representation and the position representation of $\mathfrak{osp}(1|2)$.
\end{abstract} 
\hrule
\section{Introduction}

The purpose of this paper is to obtain generating functions for the Clebsch-Gordan coefficients (CGC) of the $\mathfrak{osp}(1|2)$ Lie superalgebra. This $\mathbb{Z}_2$-graded algebra, which corresponds to the dynamical algebra of a one-dimensional para-Bose oscillator \cite{osp12}, is generated by two odd elements $J_\pm$ and one even element $J_0$. The abstract $\mathbb{Z}_2$ grading of $\mathfrak{osp}(1|2)$ can be concretized by introducing a grade involution operator $R$ ($R^2=1$) which commutes/anticommutes with the even/odd elements and by adding it to the set of generators. The $\mathfrak{osp}(1|2)$ algebra can thus be presented as the associative algebra with generators $J_0$, $J_{\pm}$, $R$ and relations
\begin{align}
  [J_0,J_\pm]=\pm J_\pm,\quad [J_0,R]=0,\quad \{J_+,J_-\}=2J_0,\quad \{J_\pm,R\}=0,\quad R^2=1,\label{relations}
\end{align}
where $[a,b]=ab-ba$ is the commutator and $\{a,b\}=ab+ba$ is the anticommutator. The presentation \eqref{relations} of $\mathfrak{osp}(1|2)$ has sometimes been referred to as $\mathfrak{sl}_{-1}(2)$, as it can be obtained from a $q\rightarrow -1$ limit of the quantum algebra $U_{q}(\mathfrak{sl}(2))$ \cite{parabosonic}. The algebra \eqref{relations} has a Casimir operator which can be written as:
\begin{align}
\label{above}
 C &=\left(J_{+}J_{-}-J_{0}+1/2\right)R.
\end{align}
The operator \eqref{above} corresponds to the sCasimir of $\mathfrak{osp}(1|2)$ multiplied by the grade involution \cite{1995_Lesniewski_JMathPhys_36_1457}. The irreducible representations of $\mathfrak{osp}(1|2)$ used in this paper are the positive-discrete series representations. These infinite-dimensional representations are labeled by two numbers $(\mu,\epsilon)$ with $\mu\geq 0$ and $\epsilon=\pm 1$. On the orthonormal basis vector $|n,\mu,\epsilon\rangle$ with $n=0,1,2,\ldots$, these representations are defined by the following actions
\begin{alignat}{2}
  \label{basis}
\begin{aligned}
 J_{0}\,\ket{n,\mu,\epsilon}&= (n + \mu + 1/2)\,\ket{n,\mu,\epsilon},&\quad R\ket{n,\mu,\epsilon} &= \epsilon\, (-1)^n\,\ket{n,\mu,\epsilon},
 \\
 J_{+}\,\ket{n,\mu,\epsilon} &= \sqrt{[n+1]_\mu}\,\ket{n+1,\mu,\epsilon},&\quad J_{-}\,\ket{n,\mu,\epsilon} &= \sqrt{[n]_\mu}\,\ket{n-1,\mu,\epsilon},
 \end{aligned}
\end{alignat}
where $[n]_\mu$ stands for the ``$\mu$-number''
\begin{align}
 [n]_\mu &= n + \mu\,(1-(-1)^n). \label{munbr}
\end{align}
The Casimir operator is a multiple of the identity:
\begin{align*}
 C\,|n,\mu,\epsilon\rangle &= -\epsilon \,\mu |n,\mu,\epsilon\rangle.
\end{align*}
The representations $(\mu,\epsilon)$ defined by \eqref{basis} correspond to the para-Bose oscillator model (see section \ref{sectionWavefunction}). In the presentation \eqref{relations}, the superalgebra $\mathfrak{osp}(1|2)$ has a coproduct $\Delta:\mathfrak{osp}(1|2)\rightarrow \mathfrak{osp}(1|2)\otimes \mathfrak{osp}(1|2)$ which has the form
\begin{align}
 \Delta(J_0) = J_0\otimes1+1\otimes J_0,\quad \Delta(J_\pm)=J_\pm\otimes R + 1 \otimes J_\pm,\quad \Delta(R)=R\otimes R. \label{coproduct}
\end{align}
The Clebsch-Gordan problem for $\mathfrak{osp}(1|2)$ can be posited as follows. Consider the tensor product representation $(\mu_1,\epsilon_1)\otimes (\mu_2,\epsilon_2)$ defined using \eqref{coproduct}. There are two natural bases for this representation space. The first one is the direct product basis with basis vectors $|n_1,\mu_1,\epsilon_1\rangle\otimes|n_2,\mu_2,\epsilon_2\rangle$. This basis, thereafter referred to as the \emph{uncoupled basis}, corresponds to the diagonalization of the operators $J_0\otimes 1$, $R\otimes 1$, $\Delta(J_0)$ and $\Delta(R)$. On these basis vectors, one has
\begin{align}
\label{uncoupled}
\begin{aligned}
 (J_0\otimes 1)|n_1,\mu_1,\epsilon_1\rangle\otimes|n_2,\mu_2,\epsilon_2\rangle&=(n_1+\mu_1+1/2)\;|n_1,\mu_1,\epsilon_1\rangle\otimes|n_2,\mu_2,\epsilon_2\rangle,
 \\
  (R\otimes 1)|n_1,\mu_1,\epsilon_1\rangle\otimes|n_2,\mu_2,\epsilon_2\rangle&=\epsilon_1 (-1)^{n_1}\;|n_1,\mu_1,\epsilon_1\rangle\otimes|n_2,\mu_2,\epsilon_2\rangle,
 \\
\Delta(J_0)|n_1,\mu_1,\epsilon_1\rangle\otimes|n_2,\mu_2,\epsilon_2\rangle&=(n_1+n_2+\mu_1+\mu_2+1)\;|n_1,\mu_1,\epsilon_1\rangle\otimes|n_2,\mu_2,\epsilon_2\rangle,
\\
\Delta(R)|n_1,\mu_1,\epsilon_1\rangle\otimes|n_2,\mu_2,\epsilon_2\rangle&=\epsilon_1\epsilon_2(-1)^{n_1+n_2}\;|n_1,\mu_1,\epsilon_1\rangle\otimes|n_2,\mu_2,\epsilon_2\rangle,
\end{aligned}
\end{align}
where $n_1,n_2\in \{0,1,2,\ldots\}$. The second basis, referred to as the \emph{coupled basis}, corresponds to the diagonalization of $\Delta(C)$, $\Delta(J_0)$ and $\Delta(R)$. On the coupled basis vectors $|n_{12},\mu_{12},\epsilon_{12}\rangle$, one has
\begin{align}
\label{coupled}
\begin{aligned}
 \Delta(C)\,|n_{12},\mu_{12},\epsilon_{12}\rangle&=-\epsilon_{12}\mu_{12}\,|n_{12},\mu_{12},\epsilon_{12}\rangle,
 \\
 \Delta(R)|n_{12},\mu_{12},\epsilon_{12}\rangle&=\epsilon_{12}(-1)^{n_{12}}|n_{12},\mu_{12},\epsilon_{12}\rangle,
 \\
 \Delta(J_0)|n_{12},\mu_{12},\epsilon_{12}\rangle&=(n_{12}+\mu_{12}+1/2)\,|n_{12},\mu_{12},\epsilon_{12}\rangle,
 \end{aligned}
\end{align}
where $n_{12}=0,1,2\ldots$. The possible values of $\mu_{12}$ and $\epsilon_{12}$ are determined by the irreducible content of the tensor product representation $(\mu_1,\epsilon_1)\otimes(\mu_2,\epsilon_2)$. It is known that one has the decomposition \cite{wavecoupled}
\begin{align*}
 (\mu_1,\epsilon_1)\otimes(\mu_2,\epsilon_2) = \bigoplus\limits_{j=0}^{\infty} (\mu_1 + \mu_2 +\frac{1}{2} + j, (-1)^j\epsilon_1\epsilon_2),
\end{align*}
provided that $\mu_1,\mu_2\geq 0$. As a consequence, the values of $\mu_{12}$ and $\epsilon_{12}$ are given by
\begin{align}\label{mu12}
 \mu_{12}= \mu_1 + \mu_2 +\frac{1}{2} + j,\quad \epsilon_{12}= (-1)^j\epsilon_1\epsilon_2 \qquad j=0,1,2,\ldots
 \end{align}
The Clebsch-Gordan problem for the $\mathfrak{osp}(1|2)$ algebra consists in finding the coefficients $C^{n_1 n_2}_{n_{12} ,j}$ that appear in the expansion of the coupled vectors in terms of those of the uncoupled basis
\begin{align}
&|n_{12},\mu_{12},\epsilon_{12}\rangle =\sum_{n_1,n_2}C^{n_1 n_2}_{n_{12},j}\; |n_1,\mu_1,\epsilon_1\rangle\otimes|n_2,\mu_2,\epsilon_2\rangle. \label{clebsch}
\end{align}
The Clebsch--Gordan coefficients $C^{n_1 n_2}_{n_{12} ,j}$  have already been determined in \cite{hahnalgebra,parabosonic}. They are given in terms of the dual $-1$ Hahn polynomials. These polynomials belong to the Bannai--Ito scheme, which comprises several families of bispectral orthogonal polynomials that correspond to $q\rightarrow -1$ limits of polynomials of the Askey scheme \cite{ 2012_Tsujimoto&Vinet&Zhedanov_AdvMath_229_2123}. The dual $-1$ Hahn polynomials are $q\rightarrow -1$ limits of the dual $q$-Hahn polynomials \cite{hahn}; they can also be obtained as a special case of the Complementary Bannai-Ito polynomials \cite{2013_Genest&Vinet&Zhedanov_SIGMA_9_18}. The dual $-1$ Hahn polynomial obey a discrete orthogonality relation on a finite lattice. In addition to satisfying the mandatory three-term recurrence relation, these polynomials are also eigenfunctions of a second-order Dunkl shift operator involving reflections; they are thus bispectral but lie outside the framework of Leonard duality.

In the present paper, we expand upon these results by deriving generating functions for these Clebsch--Gordan coefficients. We do so using two different approaches. The first approach is based on a method originally proposed by Granovskii and Zhedanov in \cite{3nj}. We generalize this technique to take into account the twisted coproduct \eqref{coproduct} of $\mathfrak{osp}(1|2)$ (see also \cite{twistedVdJ}). This method relies upon the coherent-state representation of the para-Bose oscillator. The second approach is based on the two-dimensional Dunkl oscillator model, a superintegrable system which is obtained by combining two independent one-dimensional para-Bose oscillators. Here, we use the wavefunctions of the Dunkl oscillator separated in Cartesian and polar coordinates as realizations of the basis vectors of the uncoupled and coupled bases to derive the generating functions. This method relies upon the position representation of the para-Bose oscillator.

The paper is organized as follows. In Section 2, the properties of the dual $-1$ Hahn polynomials are recalled. The generating functions for the $\mathfrak{osp}(1|2)$ CGC are derived using the first approach in section 3 and using the second method in section 4. A short conclusion follows.

\section{Dual -1 Hahn Polynomials}

This section reviews the main properties of the dual $-1$ Hahn polynomials. The connection between the $\mathfrak{ops}(1|2)$ Clebsch-Gordan coefficients and these polynomials is also provided.

The monic dual -1 Hahn polynomials, which involve two real parameters $\eta$, $\xi$ and a positive integer $N$, will be denoted by $R_n^{(-1)}(x ; \eta, \xi, N)$. They satisfy the recurrence relation
\begin{align*}
 R_{n+1}^{(-1)}(x ; \eta, \xi, N) + b_n R_n^{(-1)}(x ; \eta, \xi, N) + u_n R_{n-1}^{(-1)}(x ; \eta, \xi, N) = x R_n^{(-1)}(x ; \eta, \xi, N),
\end{align*}
where the coefficients are given by
\begin{align*}
 u_n = 4[n]_\xi[N-n+1]_\eta,\qquad b_n = 2([n]_\xi+[N-n]_\eta)-2\eta-2\xi-2N-1,
\end{align*}
with $[n]_\mu$ the $\mu$-numbers as defined in \eqref{munbr}. The dual $-1$ Hahn polynomials obey the orthogonality relation
\begin{align}
\sum_{j=0}^N w_j\;R_n^{(-1)}(y_j)\,R_m^{(-1)}(y_j)= \kappa_0\;u_1u_2...u_n\delta_{nm}, \label{orthogonality}
\end{align}
where the normalization $\kappa_0$, the grid points $y_s$ and the weights $w_{2s+q}$ for $s=0,1,\ldots, N$ and $q \in \{0,1\}$ have the expressions
\begin{align*}
 &\kappa_0(\eta,\xi,N) =
  \begin{cases}
   \displaystyle \frac{(-\eta-\xi-N)_{N/2}}{(1/2-\xi-N/2)_{N/2}}, & \text{$N$ even},
    \\
   \displaystyle \frac{(\eta+\xi+1)_{(N+1)/2}}{(\eta+1/2)_{(N+1)/2}}, & \text{$N$ odd},
  \end{cases}\\
 &y_{s}(\eta,\xi,N)=
 \begin{cases}
  \displaystyle(-1)^s (-2\eta-2\xi-2N+2s-1), & \text{$N$ even},
  \\
  \displaystyle(-1)^s (2\eta+2\xi+2s+1), & \text{$N$ odd},
 \end{cases}\\
 &w_{2s+q}(\eta,\xi,N) =
  \begin{cases}
   (-1)^s \displaystyle\frac{(-N/2)_{s+q}}{s!}\frac{(1/2-\eta-N/2)_s(-\eta-\xi-N)_s}{(1/2-\xi-N/2)_s (-\eta-\xi-N/2)_{s+q}}, &\text{$N$ even},\\
   (-1)^s \displaystyle\frac{(-(N-1)/2)_s}{s!}\frac{(\xi+1/2)_{s+q}(\eta+\xi+1)_s}{(\eta+1/2)_{s+q} (\eta+\xi+N/2+3/2)_s}, &\text{$N$ odd},
  \end{cases}
\end{align*}
with $(a)_n = (a)(a+1)\ldots(a+n-1)$.

The dual $-1$ Hahn polynomials can be expressed explicitly in terms of hypergeometric functions (see \cite{2001_Andrews&Askey&Roy} for the definition). When $N$ is even,
\begin{align*}
 R_{2k}^{(-1)}(x-1) &= \gamma_n^{(0)}\;{}_3F_2 \left( \begin{matrix} -n, \quad \delta+\frac{x}{4}, \quad \delta-\frac{x}{4} \\ -\frac{N}{2}, \quad -\frac{2\eta+N-1}{2} \end{matrix} ; 1 \right),\\
 R_{2k+1}^{(-1)}(x-1) &= \gamma_n^{(1)} (x-2\eta-2\xi)\; {}_3F_2 \left( \begin{matrix} -n, \quad \delta+\frac{x}{4}, \quad \delta-\frac{x}{4} \\ 1-\frac{N}{2}, \quad -\frac{2\eta+N-1}{2} \end{matrix} ; 1 \right),
\end{align*}
with $\delta = (\eta+\xi+N)/2$ and
\begin{align*}
\gamma_n^{(0)} =16^n\left(\frac{-N}{2}\right)_n\left(\frac{1-2\eta-N}{2}\right)_n,\qquad \gamma_n^{(1)} = 16^n\left(\frac{1-N}{2}\right)_n\left(\frac{1-2\eta-N}{2}\right)_n,
\end{align*}
while for $N$ odd,
\begin{align*}
 R_{2k}^{(-1)}(x-1) &= \gamma_n^{(0)} {}_3F_2 \left( \begin{matrix} -n, \quad \delta+\frac{x}{4}, \quad \delta-\frac{x}{4} \\ -\frac{N-1}{2}, \quad \eta+1+\frac{N}{2} \end{matrix} ; 1 \right),\\
 R_{2k+1}^{(-1)}(x-1) &= \gamma_n^{(1)} (x+2\eta-2\xi) {}_3F_2 \left( \begin{matrix} -n, \quad \delta+\frac{x}{4}, \quad \delta-\frac{x}{4} \\ -\frac{N-1}{2}, \quad \eta+1+\frac{N+1}{2} \end{matrix} ; 1 \right),
\end{align*}
where $\delta=(\eta+\xi+1)/2$ and 
\begin{align*}
\gamma_n^{(0)} = 16^n \left(\frac{1-N}{2}\right)_n\left(\frac{2\eta+1}{2}\right)_n,\qquad \gamma_n^{(1)} = 16^n \left(\frac{1-N}{2}\right)_n\left(\frac{2\eta+3}{2}\right)_n.
\end{align*}

It has been shown in \cite{hahnalgebra} that the $\mathfrak{osp}(1|2)$ Clebsch-Gordan coefficients $\langle n_1, \mu_1, \epsilon_1, n_2, \mu_2, \epsilon_2 | n_{12}, \mu_{12}, \epsilon_{12} \rangle$, which form a unitary matrix by definition, are given as follows in terms of the dual -1 polynomials:
\begin{align*}
 \langle n_1, \mu_1, \epsilon_1, n_2, \mu_2, \epsilon_2 | n_{12}, \mu_{12}, \epsilon_{12} \rangle &=
  2^{-n_1}\sqrt{\frac{w_j [n_2]_{\mu_2}!}{\kappa_0 [n_1]_{\mu_1}! [n_1+n_2]_{\mu_2}!}} R_{n_1}^{(-1)}(y_j ; \mu_2, \mu_1, n_1 + n_2),
\end{align*}
where the $\mu$-factorials $[n]_\mu!$ are defined as $[n]_\mu! = [n]_\mu[n-1]_\mu[n-2]_\mu\ldots[1]_\mu$ and where $w_j=w_j(\mu_2, \mu_1, n_1 + n_2)$, $\kappa_0=\kappa_0(\mu_2, \mu_1, n_1 + n_2)$ and $y_j=y_j(\mu_2, \mu_1, n_1 + n_2)$. The relation $n_1 + n_2 = n_{12} + j$ between the basis vector labels \eqref{labelsEquation} is assumed and will be explained in section \ref{decomposition}.

\section{Algebraic approach}

We shall call algebraic the first approach to obtain the generating functions for the $\mathfrak{osp}(1|2)$ CGC. It builds on the method introduced in \cite{3nj}. In order to present it clearly, we shall start by illustrating how it applies in the case of $\mathfrak{su}(1,1)$ before treating the $\mathfrak{osp}(1|2)$ case in details.

\subsection{The case of $\mathfrak{su}(1,1)$}\label{su11}

The $\mathfrak{su}(1,1)$ algebra is given by the generators $A_\pm$ and $A_0$ with the relations
\begin{align*}
 [A_0,A_\pm]=\pm A_\pm,\quad [A_+,A_-]=-2A_0
\end{align*}
and conditions
\begin{align*}
 A_0^\dagger = A_0, \quad A_\pm^\dagger = A_\mp.
\end{align*}
The Casimir operator $Q$ of this algebra is
\begin{align*}
 Q &= J_0^2 - J_+J_- - J_0.
\end{align*}
The positive-discrete series representations are labeled by one positive real number $l$.  The orthonormal basis vectors of a given representation $(l)$ will be noted $|m,l\rangle$ with the actions of the generators and of the Casimir operator given by
\begin{align*}
 A_0|m,l\rangle &= (l + m)|m,l\rangle,\\
 A_+ |m,l\rangle &= \sqrt{(m+1)(2l + m)}|m+1,l\rangle,\\
 A_- |m,l\rangle &= \sqrt{m(2l+m-1)}|m-1,l\rangle,\\
 Q |m,l\rangle &= l(l+1) |m,l\rangle.
\end{align*}
The $\mathfrak{su}(1,1)$ algebra can be endowed with a coproduct $\Delta : \mathfrak{su}(1,1) \rightarrow \mathfrak{su}(1,1)\otimes\mathfrak{su}(1,1)$ given by
\begin{align*}
 \Delta(A_i) &= A_i \otimes 1 + 1 \otimes A_i,
\end{align*}
which defines the tensor product of the positive discrete series representations. The decomposition of the tensor product of two irreducible representations in a direct sum of irreducible representations is known to be
\begin{align*}
(l_1)\otimes(l_2) = \bigoplus\limits_{k=0}^{\infty} (l_1 + l_2 +k).
\end{align*}
There are two natural bases for the $(l_1)\otimes(l_2)$ representation space. The uncoupled one, comprises the direct product of the basis vectors of $(l_1)$ and $(l_2)$, written $|m_1,l_1,m_2,l_2\rangle = |m_1,l_1\rangle \otimes |m_2,l_2\rangle$, and diagonalizes the operators $A_0\otimes 1$ and $\Delta(A_0)$:
\begin{align}
 A_0\otimes 1 |m_1,l_1,m_2,l_2\rangle &= (m_1 + l_1)|m_1,l_1,m_2,l_2\rangle,\\
 \Delta(A_0)|m_1,l_1,m_2,l_2\rangle &= (m_1 + m_2 + l_1 + l_2)|m_1,l_1,m_2,l_2\rangle. \label{su11uncoupled}
\end{align}
The coupled basis is the union of the standard bases of the representations $(l_1 + l_2 + k)$, noted $|m_{12},l_{12}\rangle$, where $l_{12} = l_1 + l_2 + k$, and corresponds to the diagonalization of $\Delta(A_0)$ and $\Delta(Q)$:
\begin{align}
 \Delta(A_0)|m_{12},l_{12}\rangle = (m_{12}+l_{12})|m_{12},l_{12}\rangle, \quad\Delta(Q)|m_{12},l_{12}\rangle = l_{12}(l_{12}+1) |m_{12},l_{12}\rangle. \label{su11coupled}
\end{align}

The Clebsch-Gordan problem can thus be stated as finding the expansion coefficients $\langle m_1,l_1,m_2,l_2 | m_{12}, l_{12} \rangle$ of one basis in terms of the other:
\begin{align*}
 | m_{12}, l_{12} \rangle = \displaystyle\sum\limits_{m_1,m_2}\langle m_1,l_1,m_2,l_2 | m_{12}, l_{12} \rangle|m_1,l_1,m_2,l_2\rangle.
\end{align*}
By comparing the action of $\Delta(A_0)$, given by \eqref{su11uncoupled} and \eqref{su11coupled}, on the left and right of $\langle m_1,l_1,m_2,l_2 | \Delta(A_0) | m_{12}, l_{12} \rangle$, we observe that these coefficients will be non-zero only if $m_1 + m_2 = m_{12} + k$.

The method we wish to use to find the CGC goes as follows in this case. An operator $X$ is chosen in the algebra and a (dual) basis is constructed from its left eigenvectors $\langle x,l |$
\begin{align*}
\langle x , l | X &= x \langle x , l |.
\end{align*}
Taking $X = A_+$, the overlap given by $\langle x,l | m,l \rangle$ can be factorized in terms of the "vacuum" or ground state amplitude $\langle x, l | 0,l\rangle$ and a monomial in $x$. Other choices for $X$ are possible, see for example \cite{su11VdJ}. This factorization is obtained by applying $X$ to the left and to the right:
\begin{align*}
\langle x,l |X|m,l\rangle &=x\langle x,l|m,l\rangle = \sqrt{(m+1)(2l + m)}\langle x,l |m+1,l\rangle,
\end{align*}
and by using this relation recursively
\begin{align}
\langle x,l |m,l\rangle = \langle x,l|0,l\rangle P_m^l(x), \quad P_m^l(x) =\frac{x^m}{\sqrt{m! (2 l )_m}}. \label{uncoupledFactorization}
\end{align}

Given the coproduct $\Delta(X) = X \otimes 1 +  1 \otimes X$, $\langle x,l_1,y,l_2 | = \langle x,l_1 |\otimes\langle y,l_2 |$ is a left eigenvector of $\Delta(X)$ with eigenvalue $x+y$ and the factorization property holds for the tensor product, or coupling, of two representations
\begin{align}
\langle x,l_1,y,l_2 | m_{12}, l_{12}\rangle &= \langle x,l_1,y,l_2 | 0, l_{12}\rangle P_{m_{12}}^{l_{12}} (x + y). \label{coupledFactorization}
\end{align}
We then observe that the overlap $\langle x,l_1,y,l_2 | m_{12}, l_{12}\rangle$ can be decomposed in two different ways. First, one can use the resolution of the identity in the uncoupled basis and factorize as in \eqref{uncoupledFactorization} the resulting overlaps on each subspace of the tensor product space
\begin{align*}
  \langle x,l_1,y,l_2 | m_{12}, l_{12}\rangle &= \displaystyle\sum\limits_{m_1,m_2} \langle x,l_1,y,l_2 |m_1,l_1,m_2,l_2\rangle \langle m_1,l_1,m_2,l_2 | m_{12}, l_{12} \rangle\\
&= \langle x,l_1,y,l_2 |0,l_1,0,l_2\rangle \displaystyle\sum\limits_{m_1,m_2} \langle m_1,l_1,m_2,l_2 | m_{12}, l_{12} \rangle P_{m_1}^{l_1}(x) P_{m_2}^{l_2}(y).
\end{align*}
Second, one can use the factorization property in the coupled basis \eqref{coupledFactorization} to find
\begin{align*}
\langle x,l_1,y,l_2 | m_{12}, l_{12}\rangle &= \langle x,l_1,y,l_2 | 0, l_{12}\rangle P_{m_{12}}^{l_{12}} (x + y)\\
&= P_{m_{12}}^{l_{12}} (x + y) \displaystyle\sum\limits_{m_1,m_2} \langle x,l_1,y,l_2 |m_1,l_1,m_2,l_2\rangle \langle m_1,l_1,m_2,l_2 | 0,l_{12} \rangle \\
&= \langle x,l_1,y,l_2 |0,l_1,0,l_2\rangle P_{m_{12}}^{l_{12}} (x + y)\displaystyle\sum\limits_{m_1,m_2} \langle m_1,l_1,m_2,l_2 | 0,l_{12} \rangle P_{m_1}^{l_1}(x) P_{m_2}^{l_2}(y).
\end{align*}
By cancelling the uncoupled vacuum overlap $\langle x,l_1,y,l_2 |0,l_1,0,l_2\rangle$ in the two different decompositions above, we obtain
\begin{align}
P_{m_{12}}^{l_{12}} (x + y) \displaystyle\sum\limits_{m_1,m_2} \langle m_1,l_1,m_2,l_2 | 0,l_{12} \rangle P_{m_1}^{l_1}(x) P_{m_2}^{l_2}(y) = \displaystyle\sum\limits_{m_1,m_2} \langle m_1,l_1,m_2,l_2 | m_{12}, l_{12} \rangle P_{m_1}^{l_1}(x) P_{m_2}^{l_2}(y)\label{CGsu11}
\end{align}
This is a polynomial equation involving the CGC. The generating functions are obtained by calculating the sum on the left hand side and introducing a relation between $x$ and $y$ so that the polynomials on the right hand side become monomials in a single variable. Recall the condition that $m_1 + m_2 = m_{12} + k$ for the CGC to be non-zero. It implies that the sums are constrained on the left hand side by $m_1 + m_2 = k$ and on the right hand side by $m_1 + m_2 = m_{12} + k$, given $m_{12}$ and $k$.

To sum the left hand side of \eqref{CGsu11}, we first obtain an expression for the coupled ground state CGC $\langle m_1,l_1,m_2,l_2|0,l_{12}\rangle$; this is done by solving the recurrence relation obtained by applying $\Delta(A_-)$ to the left and to the right and rearranging:
\begin{align*}
 \langle m_1+1,l_1,m_2,l_2| 0,l_{12} \rangle &= -\sqrt{\frac{(m_2+1)(2l_2+m_2)}{(m_1+1)(2l_1+m_1)}}\langle m_1,l_1,m_2+1,l_2| 0,l_{12} \rangle.
\end{align*}
Using the above equation recursively and taking into account the relation on the labels by writing $m_2 = k - m_1$ leads to
\begin{align}
 \langle m_1,l_1,k-m_1,l_2 | 0,l_{12} \rangle &= (-1)^{m_1} \sqrt{\binom{k}{m_1}\frac{(2l_2)_k}{(2l_1)_{m_1}(2l_2)_{k-m_1}}}\langle 0,l_1,k,l_2 | 0,l_{12} \rangle.\label{su11vacuumCGC}
\end{align}
The remaining undetermined factor is obtained up to a phase by using the CGC orthogonality relation. This yields
\begin{align}
 |\langle 0,l_1,k,l_2 | 0,l_{12}, \rangle|^2 &= \frac{(2l_1 + k -1)!(2l_1 + 2l_2 -2)!}{(2l_1 -1)!(2l_{12} - 2)!} \label{remainingcgc}.
\end{align}

We now return to the sum in the left hand side of \eqref{CGsu11}. Using \eqref{su11vacuumCGC} and $m_2 = k-m_1$ while omitting the term given by (\ref{remainingcgc}) and writing $m_1$ as $m$, one has
\begin{align*}
& \displaystyle\sum\limits_{m=0}^{k} (-1)^m \left(\binom{k}{m}\frac{(2l_2)_k}{(2l_1)_{m}(2l_2)_{k-m}}\right)^{\frac{1}{2}}  \frac{x^m}{\sqrt{m! (2 l_1 )_m}} \frac{y^{k-m}}{\sqrt{(k-m)! (2 l_2 )_{k-m}}}\\
  &= \frac{1}{\sqrt{k! (2l_2)_k}}\displaystyle\sum\limits_{m=0}^{k} (-1)^m \binom{k}{m}\frac{(2l_2)_k}{(2l_1)_m(2l_2)_{k-m}} x^m y^{k-m}\\
  &= \frac{y^k}{\sqrt{k! (2l_2)_k}}\displaystyle\sum\limits_{m=0}^{k} \frac{(-k)_m}{m!}\frac{(2l_2 + k -m)_m}{(2l_1)_m} (x/y)^m\\
  &= \frac{y^k}{\sqrt{k! (2l_2)_k}}\displaystyle\sum\limits_{m=0}^{k} \frac{(-k)_m(1-k-2l_2)_m}{(2l_1)_m} \frac{(-x/y)^m}{m!}
\end{align*}
One then sets $x = y^{-1}$ to recognize that
\begin{align}
\displaystyle\sum\limits_{m_1,m_2} \langle m_1,l_1,m_2,l_2 | 0,l_{12} \rangle P_{m_1}^{l_1}(x) P_{m_2}^{l_2}(y) = \frac{x^{-k}\langle 0,l_1,k,l_2 | 0,l_{12}, \rangle}{\sqrt{k! (2l_2)_k}}{}_2 F_1\left(\begin{matrix} -k ,\quad 1-k-2l_2 \\ 2l_1 \end{matrix} ; -x^2\right). \label{su11lhs}
\end{align}
Taking again $x = y^{-1}$ in the right hand side of (\ref{CGsu11}) and using $m_2 = m_{12} + k - m_1$ while relabelling $m_1$ as $m$, one obtains 
\begin{equation}
\displaystyle\sum\limits_{m_1,m_2} \langle m_1,l_1,m_2,l_2 | m_{12}, l_{12} \rangle P_{m_1}^{l_1}(x) P_{m_2}^{l_2}(y) = x^{-m_{12}-k} \displaystyle\sum\limits_{m=0}^{m_{12}+k} \frac{\langle m,l_1,m_{12}+k-m,l_2 | m_{12}, l_{12}\rangle}{\sqrt{m! (2 l_1 )_m}\sqrt{(m_{12}+k-m)! (2 l_2 )_{m_{12}+k-m}}} (x^2)^m. \label{su11rhs}
\end{equation}

Using \eqref{su11lhs} and \eqref{su11rhs} in \eqref{CGsu11}, substituting for $P_{m_{12}}^{l_{12}} (x + x^{-1})$ the expression given in \eqref{uncoupledFactorization} and posing $z = -x^2$, we obtain the generating function for the $\mathfrak{su}(1,1)$ CGC, which we recognize, up to normalization, as the generating function for the dual Hahn polynomials
\begin{align*}
 & {}_2 F_1\left(\begin{matrix} -k ,\quad 1-k-2l_2 \\ 2l_1 \end{matrix} ; z\right)\frac{(1 - z)^{m_{12}}}{\sqrt{m_{12}! (2 l_{12} )_{m_{12}}}} \frac{\langle 0,l_1,k,l_2 | 0,l_{12} \rangle}{\sqrt{k! (2l_2)_k}}\\
 &= \displaystyle\sum\limits_{m=0}^{m_{12}+k} \frac{(-1)^m \langle m,l_1,m_{12}+k-m,l_2 | m_{12}, l_{12}\rangle}{\sqrt{m! (2 l_1 )_m}\sqrt{(m_{12}+k-m)! (2 l_2 )_{m_{12}+k-m}}} z^m,
\end{align*}
where $\langle 0,l_1,k,l_2 | 0,l_{12} \rangle$ is given by \eqref{remainingcgc}.

The method only applies to algebras with untwisted coproducts. With some modifications, it can be adapted to certain algebras with twisted coproducts. We shall consider $\mathfrak{osp}(1|2)$ where as seen from (\ref{coproduct}), the twisting occurs because of the presence of an involution in the coproduct. The procedure to derive the generating function for the $\mathfrak{osp}(1|2)$ CGCs will be explained in the following sections.

\subsection{The case of $\mathfrak{osp}(1|2)$}

We now turn to $\mathfrak{osp}(1|2)$. The approach illustrated in section \ref{su11} involves two different factorizations of the overlaps between eigenvectors of an algebra element $X$ and the coupled basis vectors. This yields an equation that allows to identify a generating function for the CGC. Different choices of $X$ will lead to different factorizations, and thus, it is pertinent to choose $X$ appropriately. One such choice is the creation operator $J_+$ that has the relevant coherent states as left eigenvectors. The overlap between the coherent states and the coupled basis vectors will factorize in terms of a simple monomial and the ground state overlap, as in \eqref{uncoupledFactorization}. In light of (\ref{coproduct}), with such a choice, the coproduct of $X$ will be twisted, the factorization of the overlaps will not be immediate and will require a resolution of the identity in terms of the left eigenstates of $X$.

The presentation is organized as follows. The choice of the vectors to be used in the overlap will be explained in section \ref{Xoperator}. The factorization property and its extension to the tensor product of two representations will be developed in section \ref{factorization}. Some required identities and conditions for the CGC to be non-zero will be worked out in sections \ref{decomposition} and \ref{conditions}. Finally, the derivation of the analog of \eqref{CGsu11} for $\mathfrak{osp}(1|2)$ will be obtained in \ref{polynomialEquation} with the final results, corresponding to different parities, to be found in subsections \ref{even} and \ref{odd}.

\subsection{Signed coherent states} \label{Xoperator}

Let $X = J_+$. The (dual) coherent states in the representation $(\mu,\epsilon)$ of $\mathfrak{osp}(1|2)$ are the left-eigenvectors $\langle x,\mu,\epsilon|$ of $X$
\begin{align*}
\langle x,\mu,\epsilon|X = x\langle x,\mu,\epsilon|. 
\end{align*}
For this choice for $X$, the coherent states have simple monomials overlaps in the standard representation basis; however they do not admit a diagonal resolution of the identity \cite{residentity}. Hence, it will be simpler to work with related vectors, referred to as the signed coherent states, which are the left-eigenvectors, of $X^2=(J_+)^2$
\begin{align*}
 \langle x_\pm,\mu,\epsilon | X^2 = x^2 \langle x_\pm ,\mu,\epsilon|,
\end{align*}
and for which there exists a diagonal resolution of the identity \cite{residentity}
\begin{align}
 I &= \int d\lambda_{x_+} | x_+,\mu,\epsilon\rangle \langle x_+,\mu,\epsilon| + \int d\lambda_{x_-} | x_-,\mu,\epsilon\rangle \langle x_-,\mu,\epsilon|, \label{xpmIdentity}
\end{align}
for some measure $d\lambda_{x_\pm}$. These eigenvectors $\langle x_\pm ,\mu,\epsilon|$ of $X^2$ are expressed in terms of those of $X$ as follows
\begin{align}
 \langle x_\pm,\mu,\epsilon| &= \frac{1}{\sqrt{2}} [ \langle x,\mu,\epsilon| \pm \langle -x,\mu,\epsilon | ]. \label{x2vectors}
\end{align}
These vectors are not left-eigenvectors of the $X$, in fact, the left-action of $R$, the involution operator, and of $X$ on these vectors is
\begin{align}
 \langle x_\pm ,\mu,\epsilon| X &= x \langle x_\mp ,\mu,\epsilon| \label{X action on xpm},\\
 \langle x_\pm ,\mu,\epsilon| R &= \pm \epsilon \langle x_\pm ,\mu,\epsilon| \label{R action on xpm}.
\end{align}

\subsection{Factorization of the overlap} \label{factorization}

Consider the overlap in a representation $(\mu,\epsilon)$ of $\mathfrak{osp}(1|2)$ between the vectors $\langle x_\pm ,\mu,\epsilon|$ defined in \eqref{x2vectors} and the basis vectors $\ket{n, \mu, \epsilon}$ as given in \eqref{basis}. This overlap can be factorized in terms of a monomial in $x$ and the ground state overlap $\langle x_+, \mu, \epsilon| 0 ,\mu,\epsilon\rangle$ by considering the action of $X=J_+$ to the right and left in the expression following
\begin{align*}
 \langle x_\pm,\mu,\epsilon |X|n,\mu,\epsilon\rangle &= x\langle x_\mp,\mu,\epsilon |n,\mu,\epsilon\rangle\\
 &=\sqrt{[n+1]_\mu}\langle x_\pm,\mu,\epsilon |n+1,\mu,\epsilon\rangle.
\end{align*}
By recurrence, we see that
\begin{align}
 \langle x_\pm ,\mu,\epsilon|n,\mu,\epsilon\rangle &=\langle x_+ ,\mu,\epsilon|0,\mu,\epsilon\rangle Q_n^{\mu}(x), \label{osp12Factorization}\\
 Q_n^{\mu}(x) &= \frac{x^n}{\sqrt{[n]_\mu!}}. \label{Qpolynomial}
\end{align}
Notice that the overlap $\langle x_-, \mu, \epsilon| n ,\mu,\epsilon\rangle = 0$ for $n$ even and $\langle x_+, \mu, \epsilon| n ,\mu,\epsilon\rangle = 0$ for $n$ odd, as seen by the application of $R$ to the left and right in $\langle x_\pm, \mu, \epsilon | R | n ,\mu,\epsilon\rangle$ using \eqref{basis} and \eqref{R action on xpm}.

We now wish to extend this factorization to the tensor product of two representations. Consider the irreducible content $(\mu_{12},\epsilon_{12})$ of the representation space $(\mu_1,\epsilon_1) \otimes (\mu_2,\epsilon_2)$, with $\mu_{12}$ and $\epsilon_{12}$ given by \eqref{mu12}. The coproduct of $X=J_+$ is given by $\Delta(J_+) = J_+\otimes R + I\otimes J_+$. It can be shown \cite{residentity} that the action of $R$ on the coherent states is $\langle x,\mu,\epsilon |R = \epsilon\langle -x,\mu,\epsilon |$. This implies that the tensor product of coherent states $\langle x,\mu_1,\epsilon_1| \otimes \langle y,\mu_2,\epsilon_2| = \langle x,\mu_1,\epsilon_1,y,\mu_2,\epsilon_2|$ are not left-eigenvectors of $\Delta(X)$. To proceed, one defines the signed coherent states, or left-eigenvectors of $\Delta(X)^2$, using the analogue of \eqref{x2vectors}, that is
\begin{align*}
 \langle z_\pm,\mu_{12},\epsilon_{12}| &= \frac{1}{\sqrt{2}} [ \langle z,\mu_{12},\epsilon_{12}| \pm \langle -z,\mu_{12},\epsilon_{12} | ],
\end{align*}
where $\langle z,\mu_{12},\epsilon_{12}|$ are the coherent states of the representation $(\mu_{12},\epsilon_{12})$, defined as the left-eigenvectors of $\Delta(X)$
\begin{align*}
 \langle z,\mu_{12},\epsilon_{12}|\Delta(X) = z \langle z,\mu_{12},\epsilon_{12}|.
\end{align*}
Note that the equations corresponding to (\ref{X action on xpm}) and (\ref{R action on xpm}) still hold
\begin{align}
  \langle z_\pm,\mu_{12},\epsilon_{12}|\bigtriangleup(X) &= z\langle z_\mp,\mu_{12},\epsilon_{12}|, \label{X action on zpm}\\
  \langle z_\pm,\mu_{12},\epsilon_{12} |\bigtriangleup(R) &= \pm \epsilon_{12} \langle z_\pm,\mu_{12},\epsilon_{12}| \label{R action on zpm}.
\end{align}
It follows that the factorization property given by equation (\ref{osp12Factorization}) will stand in the coupled representation with the same polynomials as given in (\ref{Qpolynomial})
\begin{align}
  \langle z_\pm ,\mu_{12},\epsilon_{12}|n_{12},\mu_{12},\epsilon_{12}\rangle &=\langle z_+ ,\mu_{12},\epsilon_{12}|0,\mu_{12},\epsilon_{12}\rangle Q_{n_{12}}^{\mu_{12}}(z). \label{osp12coupledFactorization}
\end{align}
As per \eqref{xpmIdentity}, the resolution of the identity in terms of the tensor product of signed coherent states $\langle x_\pm,\mu_1,\epsilon_1,y_\pm,\mu_2,\epsilon_2|$, that will be required in the following, is diagonal.

Although the expression of the coupled signed coherent states $\langle z_\pm,\mu_{12},\epsilon_{12}|$ in terms of the uncoupled ones $\langle x_\pm,\mu_1,\epsilon_1, y_\pm,\mu_2,\epsilon_2|$ is not needed, we still have to establish the relation between the variables $x$, $y$ and $z$ which are not independent. This relation is easily obtained by applying $\Delta (X)^2$ to the left and right of $\langle z_\pm,\mu_{12},\epsilon_{12}| \Delta (X)^2 |x_\pm,\mu_1,\epsilon_1, y_\pm,\mu_2,\epsilon_2 \rangle$ to find
\begin{align}
 z^2 &= x^2 + y^2 . \label{xyzEquation}
\end{align}

Let us stress that we use left eigenvectors $\langle z,\mu_{12},\epsilon_{12}|$ of $\Delta (X)$ belonging to the space of the irreducible representation $(\mu_{12},\epsilon_{12})$ instead of the direct product space associated to $(\mu_1,\epsilon_1) \otimes (\mu_2,\epsilon_2)$. Those vectors are related with the tensor products of the left eigenvectors $\langle x,\mu_1,\epsilon_1,y,\mu_2,\epsilon_2|$ of $X$ via a resolution of the identity. As will be shown, the strength of the approach is that the explicit expression of the overlap $\langle z,\mu_{12},\epsilon_{12}| x,\mu_1,\epsilon_1, y,\mu_2,\epsilon_2\rangle$ need not be known. It is only required that a resolution of the identity exists in terms of the vectors $\langle x,\mu_1,\epsilon_1, y,\mu_2,\epsilon_2 |$. The use of the signed coherent states instead of the coherent states is only to facilitate the calculations.

\subsection{Coupled vectors decomposition} \label{decomposition}

We derive in this section explicit expressions for two subsets of the Clebsch-Gordan coefficients that are required to derive the generating function for the complete CGC. Specifically, an expression will be needed for the CGC denoted by $\langle n_1, \mu_1,\epsilon_1, n_2, \mu_2,\epsilon_2 | 0,\mu_{12},\epsilon_{12}\rangle$ and $\langle n_1, \mu_1,\epsilon_1, n_2, \mu_2,\epsilon_2 | 1,\mu_{12},\epsilon_{12}\rangle$. The labels in non-zero CGC are not independent. In fact, under the assumption that the overlap is non-zero, one can apply $\Delta(J_0)$ to the right and left of $\langle n_1, \mu_1,\epsilon_1, n_2, \mu_2,\epsilon_2 | \Delta(J_0) | n_{12},\mu_{12},\epsilon_{12}\rangle$ using \eqref{uncoupled}, \eqref{coupled} and \eqref{mu12} to obtain
\begin{align}
 n_1 + n_2 = n_{12} + j, \label{labelsEquation}
\end{align}
where $j$ is implicit in the parameters $\mu_{12}$ and $\epsilon_{12}$.

We now derive an expression for $\langle n_1, \mu_1,\epsilon_1, n_2, \mu_2,\epsilon_2 | 0,\mu_{12},\epsilon_{12}\rangle$. With $n_{12}=0$, we have $n_1 + n_2 = j$. This implies that we can write $n_2 = j - n_1$. By applying $\Delta(J_-)$ to the left and to the right in $\langle n_1, \mu_1,\epsilon_1, n_2, \mu_2,\epsilon_2 | \Delta(J_-) | 0,\mu_{12},\epsilon_{12}\rangle$, we obtain the following relation
\begin{align*}
 \langle n_1 + 1, \mu_1,\epsilon_1, n_2, \mu_2,\epsilon_2 | 0,\mu_{12},\epsilon_{12}\rangle = - \frac{(-1)^{n_2}}{\epsilon_2} \sqrt{\frac{[n_2 + 1]_{\mu_2}}{[n_1+1]_{\mu_1}}} \langle n_1, \mu_1,\epsilon_1, n_2+1, \mu_2,\epsilon_2 | 0,\mu_{12},\epsilon_{12}\rangle.
\end{align*}
Using the above relation recursively while making explicit the dependence between the labels in view of \eqref{labelsEquation} by relabeling $n_1$ as $n$ and writing $n_2 = j-n$ leads to
\begin{multline}
 \langle n, \mu_1,\epsilon_1, j - n, \mu_2,\epsilon_2 | 0,\mu_{12},\epsilon_{12}\rangle =\left(\frac{-1}{\epsilon_2}\right)^{n} (-1)^{nj - n(n + 1)/2}\sqrt{\frac{[j]_{\mu_2}[j-1]_{\mu_2}...[j-n +1]_{\mu_2}}{[n]_{\mu_1}!}}\label{vacuum} \\
  \times \langle 0, \mu_1,\epsilon_1, j, \mu_2,\epsilon_2 | 0,\mu_{12},\epsilon_{12}\rangle.
\end{multline}
From the orthogonality of the Clebsch-Gordan coefficients, we find
\begin{align}\label{vacuumCGC}
 |\langle 0, \mu_1,\epsilon_1, j, \mu_2,\epsilon_2 | 0,\mu_{12},\epsilon_{12}\rangle|^2 =
\begin{cases}
 \displaystyle \frac{(\frac{j}{2}+1+\mu_1+\mu_2)_{j/2}}{2^{j/2}(\frac{1}{2}+\mu_1)_{j/2}} &\text{for } j \text{ pair},\\
 {}\\
 \displaystyle \frac{(\frac{j+1}{2}+\mu_1+\mu_2)_{(j+1)/2}}{2^{(j+1)/2}(\frac{1}{2}+\mu_1)_{(j+1)/2}} &\text{for } j \text{ odd}.
\end{cases}
\end{align}

We can similarly determine $\langle n_1, \mu_1,\epsilon_1, n_2, \mu_2,\epsilon_2 | 1,\mu_{12},\epsilon_{12}\rangle$. Using \eqref{uncoupled}, \eqref{coupled} and applying $\Delta(J_+)$ to the left and right of $\langle n_1, \mu_1,\epsilon_1, n_2, \mu_2,\epsilon_2 | \Delta(J_+)  | 0,\mu_{12},\epsilon_{12}\rangle$, one is led to
\begin{multline*}
 \sqrt{[1]_{\mu_{12}}}\langle n_1, \mu_1,\epsilon_1, n_2, \mu_2,\epsilon_2 | 1,\mu_{12},\epsilon_{12}\rangle = \epsilon_2 (-1)^{n_2}\sqrt{[n_1]_{\mu_1}}\langle n_1 - 1, \mu_1,\epsilon_1, n_2, \mu_2,\epsilon_2 | 0,\mu_{12},\epsilon_{12}\rangle \\
  + \sqrt{[n_2]_{\mu_2}}\langle n_1, \mu_1,\epsilon_1, n_2 - 1, \mu_2,\epsilon_2 | 0,\mu_{12},\epsilon_{12}\rangle
\end{multline*}
Using \eqref{labelsEquation}, one now writes $n_1$ as $n$ and uses (\ref{vacuum}) to obtain
\begin{multline}\label{firstCGC}
 \langle n, \mu_1,\epsilon_1, 1+j-n, \mu_2,\epsilon_2 | 1,\mu_{12},\epsilon_{12}\rangle = \left(\frac{-1}{\epsilon_2}\right)^n \frac{(-1)^{nj - n(n+1)/2}}{\sqrt{[1]_{\mu_{12}}}} \sqrt{\frac{[j]_{\mu_2}[j-1]_{\mu_2}...[j-n+2]_{\mu_2}}{[n]_{\mu_1}!}}\\
 \times [[n]_{\mu_1} + [1+j-n]_{\mu_2}] \langle 0, \mu_1,\epsilon_1, j, \mu_2,\epsilon_2 | 0,\mu_{12},\epsilon_{12}\rangle
\end{multline}

\subsection{Selection rules}\label{conditions}

Let us now derive selection rules for relevant overlaps. Consider $\langle z_\pm ,\mu_{12},\epsilon_{12}|n_{12},\mu_{12},\epsilon_{12}\rangle$. Applying $\Delta(R)$ to the left and right within this overlap, using (\ref{R action on zpm}) and simplifying, one has
\begin{align*}
\pm \epsilon_{12} = \epsilon_{12} (-1)^{n_{12}},
\end{align*}
which implies that
\begin{align}
 &\langle z_-,\mu_{12},\epsilon_{12} | n_{12},\mu_{12},\epsilon_{12} \rangle = 0 \quad \text{for }n_{12}\text{ even} &\langle z_+,\mu_{12},\epsilon_{12} | n_{12},\mu_{12},\epsilon_{12} \rangle = 0 \quad\text{for }n_{12}\text{ odd}.\label{zeroZN12}
\end{align}
Let us examine similarly the overlap $\langle z_\pm,\mu_{12},\epsilon_{12}| x_\pm,\mu_1,\epsilon_1, y_\pm,\mu_2,\epsilon_2 \rangle$. Employing $\Delta(R)$, with the help of \eqref{R action on xpm} and \eqref{R action on zpm}, one finds the following relation
\begin{align*}
(\pm)_z (-1)^j = (\pm)_x (\pm)_y,
\end{align*}
where $(\pm)_a$ refers to the sign of the index of the variable $a$. One then has the following
\begin{align}\label{zeroZXY}
\begin{array}{c}
\left. \begin{array}{l}
 \langle z_+,\mu_{12},\epsilon_{12} | x_\pm,\mu_1,\epsilon_1, y_\pm,\mu_2,\epsilon_2 \rangle =0\\
 \langle z_-,\mu_{12},\epsilon_{12} | x_\mp,\mu_1,\epsilon_1, y_\pm,\mu_2,\epsilon_2 \rangle =0
\end{array}\right\} \quad \text{for j even},\\
\\
\left. \begin{array}{l}
 \langle z_+,\mu_{12},\epsilon_{12} | x_\mp,\mu_1,\epsilon_1, y_\pm,\mu_2,\epsilon_2 \rangle =0\\
 \langle z_-,\mu_{12},\epsilon_{12} | x_\pm,\mu_1,\epsilon_1, y_\pm,\mu_2,\epsilon_2 \rangle =0
\end{array}\right\} \quad \text{for j odd}.
\end{array}
\end{align}
These two conditions naturally lead to four cases. The reader should keep in mind, in what follows, that equation \eqref{labelsEquation} is always satisfied if we restrict ourselves to the non-zero Clebsch-Gordan coefficients.

\subsection{Generating function} \label{polynomialEquation}

With these preliminaries, we are now ready to derive an equation playing, for $\mathfrak{osp}(1|2)$, the same role as \eqref{CGsu11} in the $\mathfrak{su}(1,1)$ case. We will then bring this equation to a closed form and introduce a relation between the variables to obtain the CGC generating function.

To that end, we consider the overlap $\langle z_\pm,\mu_{12},\epsilon_{12}| n_{12},\mu_{12},\epsilon_{12}\rangle$ between the coupled basis vector $|n_{12}, \mu_{12}, \epsilon_{12}\rangle$ and the signed coherent states \eqref{x2vectors} of the representation $(\mu_{12},\epsilon_{12})$.  For ease of notation, we will take it as understood that the representation space is that of $(\mu_1,\epsilon_1)\otimes(\mu_2,\epsilon_2)$, with $\mu_{12}$ and $\epsilon_{12}$ functions of $j$ as per \eqref{mu12}. Accordingly, we will write $|n_{12},\mu_{12},\epsilon_{12}\rangle$ as $|n_{12},j\rangle$ and $|n_1,\mu_1,\epsilon_1,n_2,\mu_2,\epsilon_2\rangle$, the decoupled basis vectors \eqref{uncoupled}, as $|n_1,n_2\rangle$. Similarly, since both vectors in the overlap are defined on the same representation $(\mu_{12},\epsilon_{12})$, we will write $\langle z_\pm,\mu_{12},\epsilon_{12}|$ as simply $\langle z_\pm, j|$.

First, one can decompose the vector $|n_{12},j\rangle$ on the decoupled basis $|n_1,n_2\rangle$ using the Clebsch-Gordan decomposition. The overlap can then be expressed as
\begin{align*}
 \langle z_\pm,j| n_{12},j\rangle = \displaystyle\sum\limits_{n=0}^{n_{12}+j}\langle z_\pm,j|n,n_{12}+j-n\rangle\langle n,n_{12}+j-n| n_{12},j\rangle,
\end{align*}
where we have made use of \eqref{labelsEquation}. One then calls upon the resolution of the identity on the representation space $(\mu_1,\epsilon_1)\otimes(\mu_2,\epsilon_2)$, given in \eqref{xpmIdentity}, to obtain
\begin{align}\label{fact1}
  \langle z_\pm,j | n_{12},j \rangle = \iint d\lambda_{x_\pm} d\lambda_{y_\pm} \langle z_\pm,j | x_\pm y_\pm \rangle \displaystyle\sum\limits_{n=0}^{n_{12}+j} \langle x_\pm |n\rangle \langle y_\pm | n_{12}+j-n \rangle\langle n, n_{12}+j-n |n_{12},j\rangle,
\end{align}
where, again, the parameters $\mu_1$, $\mu_2$, $\epsilon_1$ and $\epsilon_2$ of the decoupled representation space are implicitly understood, justifying the notation $| x_{\pm} \rangle$ for the signed coherent states $| x_{\pm}, \mu, \epsilon \rangle$.

Secondly, the overlap can also be decomposed by applying first the coupled factorization \eqref{osp12coupledFactorization} and then applying the same process as above. Let
\begin{align}\label{modifiedFactorization}
 \langle z_\pm,j | n_{12},j \rangle =
\begin{cases}
 Q_{n_{12}}^{\mu_{12}}(z) \langle z_+,j | 0,j \rangle &\text{for }n_{12}\text{ even},\\
\\
 Q_{n_{12}-1}^{\mu_{12}}(z) \langle z_-,j | 1,j \rangle &\text{for }n_{12}\text{ odd},
\end{cases}
\end{align}
where the conditions \eqref{zeroZN12} were used. We can now use the resolution of the identity in terms of the uncoupled basis on the remaining overlap above
\begin{align*}
 \langle z_\pm,j | n_{12},j \rangle =
\begin{cases}
 Q_{n_{12}}^{\mu_{12}}(z) \displaystyle\sum\limits_{n=0}^{j} \langle z_+,j |n,j-n\rangle \langle n, j-n | 0,j \rangle &\text{for }n_{12}\text{ even},\\
\\
 Q_{n_{12}-1}^{\mu_{12}}(z) \displaystyle\sum\limits_{n=0}^{j+1} \langle z_-,j |n,1+j-n\rangle \langle n,1+j-n | 1,j \rangle &\text{for }n_{12}\text{ odd},
\end{cases}
\end{align*}
and finally, use the resolution of the identity \eqref{xpmIdentity} as before on the above, taking into account the conditions \eqref{zeroZN12}, to obtain, for $n_{12}$ even
\begin{align}\label{fact2a}
 \langle z_+,j | n_{12},j \rangle = Q_{n_{12}}^{\mu_{12}}(z) \iint d\lambda_{x_\pm} d\lambda_{y_\pm} \langle z_+,j | x_\pm y_\pm \rangle \displaystyle\sum\limits_{n=0}^{j} \langle x_\pm y_\pm |n,j-n\rangle \langle n, j-n | 0,j \rangle,
\end{align}
and for $n_{12}$ odd
\begin{align}\label{fact2b}
 \langle z_-,j | n_{12},j \rangle = Q_{n_{12}-1}^{\mu_{12}}(z) \iint d\lambda_{x_\pm} d\lambda_{y_\pm} \langle z_-,j | x_\pm y_\pm \rangle \displaystyle\sum\limits_{n=0}^{j+1} \langle x_\pm y_\pm |n,1+j-n\rangle \langle n,1+j-n | 1,j \rangle.
\end{align}

We now wish to compare the integrands of both decompositions of the overlap $\langle z_\pm,j | n_{12},j \rangle$, given that those equations must stand for any $z_\pm$ or $n_{12}$ and any representation parameters and that the parts of the integrand we wish to compare are polynomials of finite degrees. To do so, one must use the conditions given in \eqref{zeroZN12} and \eqref{zeroZXY} to determine, for each of the parity combinations of $n_{12}$ and $j$, the sign indices $(\pm)_i$, with $i=x,y,z$, of the coupled and decoupled signed coherent states for which the overlap $\langle z_\pm,\mu_{12},\epsilon_{12}| x_\pm,\mu_1,\epsilon_1, y_\pm,\mu_2,\epsilon_2 \rangle$ is non-zero. The remaining non-zero terms can then be equated. One then separates the sums in even and odd $n$ and factorizes the overlaps as in section \ref{factorization} in terms of the decoupled signed coherent states $\langle x_\pm y_\pm |$ to obtain polynomial expressions in the sums in \eqref{fact1}, \eqref{fact2a} and \eqref{fact2b}. We can now equate the sums coming from the first decomposition scheme \eqref{fact1} to the sums coming from the second one given by \eqref{fact2a} and \eqref{fact2b} by comparing the terms that are not summed in the integrands of these equations, requiring the factorization as applied in \eqref{modifiedFactorization}. This comparison leads to equalities for the sums themselves. Furthermore, one can combine the sums on even and odd $n$ on both side of the equalities to obtain a single equality for each of the parity combination of $n_{12}$ and $j$.

Let us clarify this process through an example with $n_{12}$ and $j$ even. With $n_{12}$ even, one can deduce, using \eqref{zeroZN12}, that it is suffices to consider the overlap with $\langle z_+,j|$ in \eqref{fact1}. Equating with \eqref{fact2a} yields
\begin{multline*}
 \iint d\lambda_{x_\pm} d\lambda_{y_\pm} \langle z_+,j | x_\pm y_\pm \rangle \displaystyle\sum\limits_{n=0}^{n_{12}+j} \langle x_\pm |n\rangle \langle y_\pm | n_{12}+j-n \rangle\langle n, n_{12}+j-n |n_{12},j\rangle =\\
 Q_{n_{12}}^{\mu_{12}}(z) \iint d\lambda_{x_\pm} d\lambda_{y_\pm} \langle z_+,j | x_\pm y_\pm \rangle \displaystyle\sum\limits_{n=0}^{j} \langle x_\pm y_\pm |n,j-n\rangle \langle n, j-n | 0,j \rangle.
\end{multline*}
In view of \eqref{zeroZXY}, with $j$ even, we know that the sign indices of $x_\pm$ and $y_\pm$ must be the same. Incidentally, as noted in section \ref{factorization}, when the sign index is $+$, only the terms with $n$ even in both sums of the above equation will be non-zero, whereas when the sign is $-$, only the terms with $n$ odd will contribute. The above equation is valid when restricted to the terms with $n$ even, or respectively, odd. For $n$ even, this leads to
\begin{multline}\label{evenTerms}
 \iint d\lambda_{x_+} d\lambda_{y_+} \langle z_+,j | x_+ y_+ \rangle \displaystyle\sum\limits_{\substack{n=0\\ n\text{ even}}}^{n_{12}+j} \langle x_+ |n\rangle \langle y_+ | n_{12}+j-n \rangle\langle n, n_{12}+j-n |n_{12},j\rangle =\\
 Q_{n_{12}}^{\mu_{12}}(z) \iint d\lambda_{x_+} d\lambda_{y_+} \langle z_+,j | x_+ y_+ \rangle \displaystyle\sum\limits_{\substack{n=0\\ n\text{ even}}}^{j} \langle x_+ y_+ |n,j-n\rangle \langle n, j-n | 0,j \rangle,
\end{multline}
It is now possible to apply the factorization of section \ref{factorization} to the uncoupled signed coherent states overlaps. For $n$ even, one obtains, from \eqref{evenTerms}
\begin{multline*}
  \iint d\lambda_{x_+} d\lambda_{y_+} \langle z_+,j | x_+ y_+ \rangle \langle x_+ | 0 \rangle \langle y_+ | 0 \rangle \displaystyle\sum\limits_{\substack{n=0\\ n\text{ even}}}^{n_{12}+j} Q_n^{\mu_1}(x) Q_{n_{12}+j-n}^{\mu_2}(y) \langle n, n_{12}+j-n |n_{12},j\rangle =\\
  \iint d\lambda_{x_+} d\lambda_{y_+} \langle z_+,j | x_+ y_+ \rangle \langle x_+ | 0 \rangle \langle y_+ | 0 \rangle Q_{n_{12}}^{\mu_{12}}(z) \displaystyle\sum\limits_{\substack{n=0\\ n\text{ even}}}^{j} Q_n^{\mu_1}(x) Q_{j-n}^{\mu_2}(y) \langle n, j-n | 0,j \rangle.
\end{multline*}
As explained before, we can now equate the integrands, leading to
\begin{align*}
 \displaystyle\sum\limits_{\substack{n=0\\ n\text{ even}}}^{n_{12}+j} Q_n^{\mu_1}(x) Q_{n_{12}+j-n}^{\mu_2}(y) \langle n, n_{12}+j-n |n_{12},j\rangle = Q_{n_{12}}^{\mu_{12}}(z) \displaystyle\sum\limits_{\substack{n=0\\ n\text{ even}}}^{j} Q_n^{\mu_1}(x) Q_{j-n}^{\mu_2}(y) \langle n, j-n | 0,j \rangle
\end{align*}
For $n$ odd, instead of \eqref{evenTerms}, one gets
\begin{align*}
 \displaystyle\sum\limits_{\substack{n=1\\ n\text{ odd}}}^{n_{12}+j-1} Q_n^{\mu_1}(x) Q_{n_{12}+j-n}^{\mu_2}(y) \langle n, n_{12}+j-n |n_{12},j\rangle = Q_{n_{12}}^{\mu_{12}}(z) \displaystyle\sum\limits_{\substack{n=1\\ n\text{ odd}}}^{j-1} Q_n^{\mu_1}(x) Q_{j-n}^{\mu_2}(y) \langle n, j-n | 0,j \rangle.
\end{align*}
The two equations above can be combined to give
\begin{align}\label{purepolynomialEquation}
 \displaystyle\sum\limits_{n=0}^{n_{12}+j} Q_n^{\mu_1}(x) Q_{n_{12}+j-n}^{\mu_2}(y) \langle n, n_{12}+j-n |n_{12},j\rangle = Q_{n_{12}}^{\mu_{12}}(z) \displaystyle\sum\limits_{n=0}^{j} Q_n^{\mu_1}(x) Q_{j-n}^{\mu_2}(y) \langle n, j-n | 0,j \rangle.
\end{align}

Equations analogous to \eqref{purepolynomialEquation} are similarly obtained for each parity combination for $n_{12}$ and $j$. In order to obtain generating functions, we wish to bring the right hand side of \eqref{purepolynomialEquation}, or its equivalent in other parity cases, to a closed form while introducing a relation between the variables $x$ and $y$ to transform the left hand side sum in a power series of a single variable. This work is best done treating the two parities of $n_{12}$ separately and those of $j$ in parallel.

\subsubsection{Cases with $n_{12}$ even} \label{even}

We have derived above \eqref{purepolynomialEquation} an equation playing for $\mathfrak{osp}(1|2)$ the same role that \eqref{CGsu11} plays in the case of $\mathfrak{su}(1,1)$. This equation is valid for $n_{12}$ and $j$ even. It can easily be shown that the same expression is obtained when one considers the case with $n_{12}$ even and $j$ odd. We will thus construct the generating function for both parities of $j$ in parallel. Using \eqref{Qpolynomial} and \eqref{vacuum} in \eqref{purepolynomialEquation} and simplifying, we obtain
\begin{multline}\label{steptoGenFunEven}
  y^{n_{12}}\displaystyle\sum\limits_{n=0}^{n_{12}+j} \frac{(x/y)^n}{\sqrt{[n]_{\mu_1}!}}  \frac{\langle n, n_{12}+j-n |n_{12},j\rangle}{\sqrt{[n_{12}+j-n]_{\mu_2}!}} = \displaystyle\sum\limits_{n=0}^{j} \left(\frac{-x}{\epsilon_2 y}\right)^{n} (-1)^{nj - n(n + 1)/2} \frac{[j]_{\mu_2}[j-1]_{\mu_2}...[j-n +1]_{\mu_2}}{[n]_{\mu_1}!}\\
  \times \frac{z^{n_{12}}}{\sqrt{[n_{12}]_{\mu_{12}}![j]_{\mu_2}!}} \langle0,j|0,j\rangle,
\end{multline}
where $\langle0,j|0,j\rangle = \langle 0, \mu_1,\epsilon_1, j, \mu_2,\epsilon_2 | 0,\mu_{12},\epsilon_{12}\rangle$ is given by \eqref{vacuumCGC} and where we made use of the following identity
\begin{align}\label{muFacttoPartialmuFact}
 \frac{1}{[j-n]_\mu !} = \frac{[j]_{\mu}[j-1]_{\mu}...[j-n +1]_{\mu}}{[j]_\mu !}.
\end{align}
We now want to reduce the left hand side of \eqref{steptoGenFunEven} to the form of a generating function by writing it as a power series of a single variable. Setting $y=x^{-1}$ leads to
\begin{multline}\label{dirtyGenFunEven}
   \displaystyle\sum\limits_{n=0}^{n_{12}+j} \frac{x^{2n}}{\sqrt{[n]_{\mu_1}!}}  \frac{\langle n, n_{12}+j-n |n_{12},j\rangle}{\sqrt{[n_{12}+j-n]_{\mu_2}!}} = \displaystyle\sum\limits_{n=0}^{j} \left(\frac{-x^2}{\epsilon_2}\right)^{n} (-1)^{nj - n(n + 1)/2} \frac{[j]_{\mu_2}[j-1]_{\mu_2}...[j-n +1]_{\mu_2}}{[n]_{\mu_1}!}\\
  \times \frac{(xz)^{n_{12}}}{\sqrt{[n_{12}]_{\mu_{12}}![j]_{\mu_2}!}} \langle0,j|0,j\rangle.
\end{multline}
In order to express the right hand side of \eqref{dirtyGenFunEven} in a closed form, one needs to split the terms with even and odd $n$. This separation enables one to write the $\mu$-factorials in terms of Pochhammer symbols. We will need the following identities
\begin{align}\label{muFacttoPoch}
 [n]_{\mu}! &=
 \begin{cases}
  2^n k! (\frac{1}{2} + \mu)_k &\text{ for }n\text{ even, }n=2k,\\
  2^n k! (\frac{3}{2} + \mu)_k (\frac{1}{2} + \mu) &\text{ for }n\text{ odd, }n=2k+1,
 \end{cases}
\end{align}
\begin{align}\label{partialmuFacttoPoch}
[j]_{\mu}[j-1]_{\mu}...[j-n +1]_{\mu} =
\begin{cases}
 2^n (-\frac{j}{2})_k (\frac{1}{2}-\frac{j}{2}-\mu)_k &\text{ for }n,j\text{ even, }n=2k,\\
 2^n (1-\frac{j}{2})_k (\frac{1}{2}-\frac{j}{2}-\mu)_k (\frac{j}{2}) &\text{ for }n\text{ odd and }j\text{ even, }n=2k+1,\\
 2^n (\frac{1}{2}-\frac{j}{2})_k (-\frac{j}{2}-\mu)_k &\text{ for }n\text{ even and }j\text{ odd, }n=2k,\\
 2^n (1-\frac{j}{2}-\mu)_k (\frac{1}{2}-\frac{j}{2})_k &\text{ for }n,j\text{ odd, }n=2k+1,
\end{cases}
\end{align}
\begin{align}\label{signSimplicfication}
 (-1)^{nj - n(n + 1)/2} =
 \begin{cases}
  (-1)^k &\text{for }n\text{ even, }n=2k,\\
  (-1)^{j+k+1} &\text{for }n\text{ odd, }n=2k+1.
 \end{cases}
\end{align}

We can now derive the final result. Separating the sums in even and odd $n$ in the right hand side of \eqref{dirtyGenFunEven}, using \eqref{muFacttoPoch}, \eqref{partialmuFacttoPoch} and \eqref{signSimplicfication} while changing the variable to $s=x^2$, one obtains that the sum on the right hand side of \eqref{dirtyGenFunEven} can be written in terms of hypergeometric functions. That is,
\begin{multline}\label{hypergeoSumEven}
\displaystyle\sum\limits_{n=0}^{j} \left(\frac{-x^2}{\epsilon_2}\right)^{n} (-1)^{nj - n(n + 1)/2} \frac{[j]_{\mu_2}[j-1]_{\mu_2}...[j-n +1]_{\mu_2}}{[n]_{\mu_1}!}\\
=
\begin{cases}
{}_2 F_1\left(\begin{matrix} -\frac{j}{2},\quad \frac{1}{2}-\frac{j}{2}-\mu_2 \\ \frac{1}{2} + \mu_1 \end{matrix} ; -s^2\right) +  \displaystyle\frac{js}{(1 + 2\mu_1) \epsilon_2} {}_2 F_1\left(\begin{matrix}1-\frac{j}{2},\quad \frac{1}{2}-\frac{j}{2}-\mu_2 \\ \frac{3}{2} + \mu_1 \end{matrix} ; -s^2\right) &\text{for }j\text{ even},\\
{}\\
{}_2 F_1\left(\begin{matrix} \frac{1}{2}-\frac{j}{2},\quad -\frac{j}{2}-\mu_2 \\ \frac{1}{2} + \mu_1 \end{matrix} ; -s^2\right) - \displaystyle s\frac{(j + 2\mu_2)}{(1 + 2\mu_1) \epsilon_2} {}_2 F_1\left(\begin{matrix} 1-\frac{j}{2}-\mu_2,\quad \frac{1}{2}-\frac{j}{2} \\ \frac{3}{2} + \mu_1 \end{matrix} ; -s^2\right) &\text{for }j\text{ odd}.
\end{cases}
\end{multline}

The remaining term in \eqref{dirtyGenFunEven} is still expressed as a function of $z$ and not $s$. Using \eqref{xyzEquation}, and $y=x^{-1}$, one easily finds
\begin{align}\label{sqrtTermEven}
 \frac{(xz)^{n_{12}}}{\sqrt{[n_{12}]_{\mu_{12}}![j]_{\mu_2}!}} \langle0,j|0,j\rangle = \frac{(s^2 + 1)^{n_{12}/2}}{\sqrt{[j]_{\mu_2}! [n_{12}]_{\mu_{12}}!}}\langle0,j|0,j\rangle.
\end{align}
When $n_{12}$ is even, using \eqref{hypergeoSumEven} and \eqref{sqrtTermEven} in \eqref{dirtyGenFunEven} with $s=x^2$, the generating function for the $\mathfrak{osp}(1|2)$ Clebsch-Gordan coefficients is found to be:
\begin{multline*}
 \displaystyle\sum\limits_{n=0}^{n_{12}+j} \frac{s^n}{\sqrt{[n]_{\mu_1}!}}  \frac{\langle n, n_{12}+j-n |n_{12},j\rangle}{\sqrt{[n_{12}+j-n]_{\mu_2}!}} = \frac{(s^2 + 1)^{n_{12}/2}}{\sqrt{[j]_{\mu_2}! [n_{12}]_{\mu_{12}}!}}\langle0,j|0,j\rangle\\
\times
\begin{cases}
{}_2 F_1\left(\begin{matrix} -\frac{j}{2},\quad \frac{1}{2}-\frac{j}{2}-\mu_2 \\ \frac{1}{2} + \mu_1 \end{matrix} ; -s^2\right) +  \displaystyle\frac{js}{(1 + 2\mu_1) \epsilon_2} {}_2 F_1\left(\begin{matrix}1-\frac{j}{2},\quad \frac{1}{2}-\frac{j}{2}-\mu_2 \\ \frac{3}{2} + \mu_1 \end{matrix} ; -s^2\right) &\text{for }j\text{ even},\\
{}\\
{}_2 F_1\left(\begin{matrix} -\frac{j-1}{2},\quad -\frac{j}{2}-\mu_2 \\ \frac{1}{2} + \mu_1 \end{matrix} ; -s^2\right) - \displaystyle s\frac{(j + 2\mu_2)}{(1 + 2\mu_1) \epsilon_2} {}_2 F_1\left(\begin{matrix} -\frac{j-1}{2},\quad 1-\frac{j}{2}-\mu_2 \\ \frac{3}{2} + \mu_1 \end{matrix} ; -s^2\right) &\text{for }j\text{ odd},
\end{cases}
\end{multline*}
where again, $\langle0,j|0,j\rangle = \langle 0, \mu_1,\epsilon_1, j, \mu_2,\epsilon_2 | 0,\mu_{12},\epsilon_{12}\rangle$ is given by \eqref{vacuumCGC}.

\subsubsection{Cases with $n_{12}$ odd} \label{odd}

Starting from \eqref{fact1} but using now \eqref{fact2b} as factorization, one can derive an equation equivalent to \eqref{purepolynomialEquation} in the case with $n_{12}$ odd. Again, this equation has the same expression for both parities of $j$. The analogue of \eqref{CGsu11} in the case of $\mathfrak{osp}(1|2)$ with $n_{12}$ odd is thus
\begin{align*}
 \displaystyle\sum\limits_{n = 0}^{n_{12} + j}  Q_{n}^{\mu_1}(x) Q_{n_{12}+j-n}^{\mu_2}(y) \langle n, n_{12}+j-n |n_{12},j\rangle = \displaystyle\sum\limits_{n = 0}^{j+1}  Q_{n}^{\mu_1}(x) Q_{1+ j-n}^{\mu_2}(y) \langle n, 1+j-n | 1,j\rangle Q_{n_{12}-1}^{\mu_{12}}(z)
\end{align*}
Using \eqref{Qpolynomial}, \eqref{xyzEquation}, \eqref{firstCGC} and \eqref{muFacttoPartialmuFact} in the above, simplifying and introducing the relation $y=x^{-1}$ on the variables leads to
\begin{multline}\label{dirtyGenFunOdd}
 \displaystyle\sum\limits_{n = 0}^{n_{12} + j}   \frac{x^{2n}}{\sqrt{[n]_{\mu_1}!}} \frac{\langle n, n_{12}+j-n |n_{12},j\rangle}{\sqrt{[n_{12}+j-n]_{\mu_2}!}} = \displaystyle\sum\limits_{n = 0}^{j+1}  \left(\frac{-x^2}{\epsilon_2}\right)^n (-1)^{nj - n(n+1)/2} \frac{[j]_{\mu_2}[j-1]_{\mu_2}...[j-n+2]_{\mu_2}}{[n]_{\mu_1}!}\\
 \times [[n]_{\mu_1} + [1+j-n]_{\mu_2}] \frac{(x^4+1)^{(n_{12}-1)/2}}{\sqrt{[n_{12}-1]_{\mu_{12}}![1]_{\mu_{12}}}} \frac{\langle 0, j | 0,j\rangle}{\sqrt{[j]_{\mu_2}!}}
\end{multline}
To express the right hand side of \eqref{dirtyGenFunOdd} in a closed form, the following identity are needed
\begin{multline*}
([n]_{\mu_1} + [1+j-n]_{\mu_2})[j]_{\mu_2}[j-1]_{\mu_2}...[j-n+2]_{\mu_2}\\
=
\begin{cases}
 2^n (-\frac{j}{2})_k (-\frac{j+1}{2}-\mu_2)_k &\text{ for }n,j\text{ even, }n=2k,\\
 2^n (-\frac{j}{2})_k (-\frac{j-1}{2}-\mu_2)_k (j+1+2\mu_1)/2 &\text{ for }n\text{ odd and }j\text{ even, }n=2k+1,\\
 2^n (-\frac{j+1}{2})_k (-\frac{j}{2}-\mu_2)_k &\text{ for }n\text{ even and }j\text{ odd, }n=2k,\\
 2^n (-\frac{j-1}{2})_k (-\frac{j}{2}-\mu_2)_k (1+j+2\mu_1 +2\mu_2)/2 &\text{ for }n,j\text{ odd, }n=2k+1.
\end{cases}
\end{multline*}
Using these, \eqref{muFacttoPoch} and \eqref{signSimplicfication} when splitting the right hand side of \eqref{dirtyGenFunOdd} in two sums on the even and odd $n$ leads to an expression in terms of hypergeometric series. Taking $s=x^2$, this expression is given by
\begin{multline}\label{hypergeoSumOdd}
 \displaystyle\sum\limits_{n = 0}^{j+1}  \left(\frac{-x^2}{\epsilon_2}\right)^n (-1)^{nj - n(n+1)/2} \frac{[j]_{\mu_2}[j-1]_{\mu_2}...[j-n+2]_{\mu_2}}{[n]_{\mu_1}!}[[n]_{\mu_1} + [1+j-n]_{\mu_2}]\\
=
\begin{cases}
 {}_2 F_1\left(\begin{matrix} -\frac{j}{2},\quad -\frac{j+1}{2}-\mu_2 \\ \frac{1}{2} + \mu_1 \end{matrix} ; -s^2\right) + \displaystyle s\frac{(j+1+2\mu_1)}{(1 + 2\mu_1) \epsilon_2} {}_2 F_1\left(\begin{matrix} -\frac{j}{2},\quad -\frac{j-1}{2}-\mu_2 \\ \frac{3}{2} + \mu_1 \end{matrix} ; -s^2\right) &\text{for }j\text{ even},\\
 {}\\
 {}_2 F_1\left(\begin{matrix} -\frac{j+1}{2},\quad -\frac{j}{2}-\mu_2 \\ \frac{1}{2} + \mu_1 \end{matrix} ; -s^2\right) - \displaystyle s\frac{(1+j+2\mu_1 +2\mu_2)}{(1 + 2\mu_1) \epsilon_2} {}_2 F_1\left(\begin{matrix} -\frac{j-1}{2},\quad -\frac{j}{2}-\mu_2 \\ \frac{3}{2} + \mu_1 \end{matrix} ; -s^2\right) &\text{for }j\text{ even}.
\end{cases}
\end{multline}
One can then use \eqref{hypergeoSumOdd} in \eqref{dirtyGenFunOdd} with $s=x^2$, to obtain the generating function for the $\mathfrak{osp}(1|2)$ Clebsch-Gordan coefficients when $n_{12}$ is odd
\begin{multline*}
 \displaystyle\sum\limits_{n = 0}^{n_{12} + j}   \frac{s^n}{\sqrt{[n]_{\mu_1}!}} \frac{\langle n, n_{12}+j-n |n_{12},j\rangle}{\sqrt{[n_{12}+j-n]_{\mu_2}!}} = \frac{(s^2+1)^{(n_{12}-1)/2}}{\sqrt{[n_{12}-1]_{\mu_{12}}![1]_{\mu_{12}}}} \frac{\langle 0, j | 0,j\rangle}{\sqrt{[j]_{\mu_2}!}}\\
\times
\begin{cases}
 {}_2 F_1\left(\begin{matrix} -\frac{j}{2},\quad -\frac{j+1}{2}-\mu_2 \\ \frac{1}{2} + \mu_1 \end{matrix} ; -s^2\right) + \displaystyle s\frac{(j+1+2\mu_1)}{(1 + 2\mu_1) \epsilon_2} {}_2 F_1\left(\begin{matrix} -\frac{j}{2},\quad -\frac{j-1}{2}-\mu_2 \\ \frac{3}{2} + \mu_1 \end{matrix} ; -s^2\right) &\text{for }j\text{ even},\\
 {}\\
 {}_2 F_1\left(\begin{matrix} -\frac{j+1}{2},\quad -\frac{j}{2}-\mu_2 \\ \frac{1}{2} + \mu_1 \end{matrix} ; -s^2\right) - \displaystyle s\frac{(1+j+2\mu_1 +2\mu_2)}{(1 + 2\mu_1) \epsilon_2} {}_2 F_1\left(\begin{matrix} -\frac{j-1}{2},\quad -\frac{j}{2}-\mu_2 \\ \frac{3}{2} + \mu_1 \end{matrix} ; -s^2\right) &\text{for }j\text{ even},
\end{cases}
\end{multline*}
where we still have that $\langle0,j|0,j\rangle = \langle 0, \mu_1,\epsilon_1, j, \mu_2,\epsilon_2 | 0,\mu_{12},\epsilon_{12}\rangle$ is given by \eqref{vacuumCGC}.

\section{Wavefunction Approach}\label{sectionWavefunction}

It is known that the $\mathfrak{osp}(1|2)$ algebra is the dynamical algebra of a parabosonic oscillator \cite{parabosonic} and that it admits a realization in terms of differential-difference operators acting on wavefunctions of this system. Consider then two independent parabosonic oscillators in the variables $x$ and $y$. One can identify these variables as the Cartesian coordinates in the plane. The Schrödinger equation for this two-dimensional system separates naturally in these Cartesian coordinates, but also in polar coordinates. The associated two sets of wavefunctions each constitute a basis of the Hilbert space associated with the problem. This implies that they can be expanded one in term of the other. The coefficients of this expansion will be shown to be the Clebsch-Gordan coefficients of $\mathfrak{osp}(1|2)$. The generating functions of the CGC can also be derived in this framework. We will assume, in what follows, that $\epsilon=1$ and shall omit this parameter in the notation from now on and work with $(\mu,1)$ representations.

The realization of the presentation \eqref{relations} of $\mathfrak{osp}(1|2)$ as a dynamical algebra goes as follows. Let $\mathfrak{D}_x$ be the Dunkl derivative defined by
\begin{align*}
 \mathfrak{D}_x &= \partial_x + \frac{\mu}{x} (1-P_x),
\end{align*}
where the parity operator $P_x$ acts on functions according to $P_x f(x) = f(-x)$. It is straightforward to verify that the operators
\begin{align*}
 J_0 &= -\frac{1}{2}\mathfrak{D}_x^2 + \frac{1}{2} x^2, \quad J_\pm = \frac{1}{\sqrt{2}}(x\mp \mathfrak{D}_x), \quad R = P_x,
\end{align*}
realize the $\mathfrak{osp}(1|2)$ superalgebra. The operator $J_0$ is then identified as the parabose Hamiltonian $H$. Thus, the associated Hilbert space is spanned by $\mathfrak{osp}(1|2)$ representation basis vectors $|n,\mu\rangle$ given in \eqref{basis}, viewed as solutions to the eigenvalue equation $H |\psi\rangle = E |\psi\rangle$ with $E=n+\mu+1/2$. One can identify the position operator as
\begin{align*}
 &X = \frac{1}{\sqrt{2}} (J_+ + J_-), \quad \langle x | X = x \langle x|.
\end{align*}
The wavefunctions are the overlaps between the left-eigenvectors of $X$ and the basis vectors $|n,\mu\rangle$, explicitly
\begin{align}\label{wavefunctionsAbstract}
 \psi_n^\mu(x) = \langle x | n,\mu \rangle,\quad H \psi_n^\mu(x) = (n + \mu + 1/2)\psi_n^\mu(x).
\end{align}

Consider now two independent one-dimensional parabosonic oscillators in the variables $x$ and $y$ and let their respective Hamiltonian be $H_x$ and $H_y$. By identifying these variables as Cartesian coordinates in the plane we can construct a two-dimensional problem with Hamiltonian $H_{xy} = H_x + H_y$ acting on the tensor product of two $\mathfrak{osp}(1|2)$ representations $(\mu_1,1) \otimes (\mu_2,1)$. The product of the respective wavefunctions form a basis of the Hilbert space of the two-dimensional problem as solutions of the eigenvalue equation for the total Hamiltonian $H_{xy}$:
\begin{align*}
H_{xy} \psi_{n_1}^{\mu_1}(x)\psi_{n_2}^{\mu_2}(y) = (n_1 + n_2 + \mu_1 + \mu_2 + 1) \psi_{n_1}^{\mu_1}(x)\psi_{n_2}^{\mu_2}(y).
\end{align*}
These bivariate wavefunctions also diagonalize $H_x$, $H_y$ and the realizations $C_x$ and $C_y$ of the Casimir operators \eqref{above} and we will refer to them as the \emph{uncoupled wavefunctions}.

The basis vectors $|n_{12},\mu_{12},\epsilon_{12}\rangle$ given in \eqref{coupled} also diagonalize the Hamiltonian $H_{xy}$ of the two-dimensional problem since $H_{xy} = J_0\otimes 1 + 1 \otimes J_0 = \Delta(J_0)$. One can thus build wavefunctions by considering the overlap between the tensor product $\langle x,y| = \langle x |\otimes \langle y|$ of two left-eigenvectors of $X$ and the coupled basis vectors
\begin{align}
 \Psi_{n_{12}}^{\mu_{12},\epsilon_{12}}(x,y) &= \langle x,y|n_{12},\mu_{12},\epsilon_{12}\rangle,\\
 H_{xy} \Psi_{n_{12}}^{\mu_{12},\epsilon_{12}}(x,y) &= (n_{12} + \mu_{12} + 1/2)\Psi_{n_{12}}^{\mu_{12},\epsilon_{12}}(x,y).
\end{align}
These wavefunctions also diagonalize the realization of the Casimir operator $C_{xy}$ of $(\mu_{12},\epsilon_{12})$ given by
\begin{align}\label{CasimirRealization}
 C_{xy} &= (y\mathfrak{D}_x - x\mathfrak{D}_y)P_x -\mu_1 P_y - \mu_2 P_x - (1/2) P_x P_y,
\end{align}
and will thus be called the \emph{coupled wavefunctions}. Since the union of the vectors $|n_{12},\mu_{12},\epsilon_{12}\rangle$ for each irreducible $(\mu_{12},\epsilon_{12})$ contained in the tensor product $(\mu_1,1)\otimes(\mu_2,1)$ is a basis of the total tensor product space, the wavefunctions thus obtained form a complete basis of the two-dimensional Hilbert space.

One can relate these coupled wavefunctions to the separation in polar coordinates since the Casimir operator \eqref{CasimirRealization} they diagonalize is the generalized Dunkl angular momentum associated with this separation (see \cite{wavecoupled}). Using $x = \rho \cos \phi$ and $y = \rho \sin \phi$, we will thus write the coupled wavefunctions as
\begin{align*}
 \Psi_{n_{12}}^{\mu_{12},\epsilon_{12}}(\rho,\phi) = P_{n_{12}}^{\mu_{12}}(\rho)F_j^{\mu_1,\mu_2}(\phi),
\end{align*}
where $j$ is implicitly given by \eqref{mu12} and where the representation parameters $\mu_1$ and $\mu_2$ are considered fixed for two given parabosonic oscillators associated with the representation space $(\mu_1,1) \otimes (\mu_2,1)$.

The coupled and uncoupled wavefunctions being a base of the same Hilbert space, they must decompose onto each other. Consider the eigenspace spanned by $\Psi_{n_{12}}^{\mu_{12},\epsilon_{12}}(\rho,\phi)$ associated to the eigenvalue $n_{12} + \mu_{12} + 1/2$ of $H_{xy}$. The same eigenspace is spanned by a linear combination of the uncoupled wavefunctions with eigenvalues $n_1 + n_2 + \mu_1 + \mu_2 +1$. It is easy to see that the coefficients in this combination will be the CGC:
\begin{align*}
 \langle x,y|n_{12},\mu_{12},\epsilon_{12} \rangle &= \displaystyle\sum\limits_{n1,n2} \langle x | n_1,\mu_1\rangle \langle y | n_2,\mu_2\rangle \langle n_1, n_2, \mu_1, \mu_2 |n_{12},\mu_{12},\epsilon_{12}\rangle.
\end{align*}
By comparing the eigenvalues of $H_{xy}$ on both the coupled and uncoupled wavefunctions, one can see, that the uncoupled wavefunctions will only overlap the eigenspace spanned by $\Psi_{n_{12}}^{\mu_{12},\epsilon_{12}}(\rho,\phi)$ if $n_{12} + j = n_1 + n_2$, as expected. These observations allow one to derive the $\mathfrak{osp}(1|2)$ Clebsch-Gordan coefficients generating function from analytical manipulations on the functional relations obtained in this context.

\subsection{Decoupled Wavefunctions}

The eigenfunctions of the one-dimensional parabose oscillator Hamiltonian are \cite{wavedecoupled}
\begin{align}\label{CartesianWavefunctions}
 \psi_n^\mu(x) &= e^{\frac{-x^2}{2}} H_n^{\mu}(x),
\end{align}
where $H_n^{\mu}(x)$ are the generalized Hermite polynomials
\begin{align}\label{generalizedHermite}
H_{2k+p}^{\mu}(x) &= (-1)^k \sqrt{\frac{k!}{\Gamma(k+p+\mu+1/2)}}x^p L_k^{\mu-1/2+p}(x^2),
\end{align}
with $p \in \{0,1\}$ and $k=0,1,2,...$ and where $L_n^{\alpha}(x)$ are Laguerre polynomials. The decoupled wavefunctions for the two-dimensional system are then given by $\psi_{n_1}^{\mu_1}(x)\psi_{n_2}^{\mu_2}(y)$, with eigenvalue $n_1 + n_2 + \mu_1 + \mu_2 + 1$.

\subsection{Coupled Wavefunctions}\label{coupledWavefunctions}

Solving the simultaneous eigenvalue equations for the Hamiltonian $H_{xy}$ and the total Casimir $C_{xy}$, one obtains \cite{wavecoupled}
\begin{align}\label{CoupledAsRadialAngular}
 \Psi_{n_{12}}^{\mu_{12},\epsilon_{12}}(\rho,\phi) &= P_{n_{12}}^{\mu_{12}}(\rho)F_j^{\mu_1,\mu_2}(\phi),
\end{align}
where the radial wavefunctions are given by the following
\begin{align*}
P_{n_{12}}^{\mu_{12}}(\rho)=
\begin{cases}
 \displaystyle\sqrt{\frac{2\frac{n_{12}}{2}!}{\Gamma(\frac{n_{12}}{2}+j+\mu_1+\mu_2+1)}} e^{-\rho^2/2}\rho^{j}L_{\frac{n_{12}}{2}}^{j+\mu_1+\mu_2}(\rho^2) &\text{ for }n_{12},j\text{ even,}\\
 \displaystyle\sqrt{\frac{2\frac{n_{12}-1}{2}!}{\Gamma(\frac{n_{12}-1}{2}+j+\mu_1+\mu_2+2)}} e^{-\rho^2/2}\rho^{j+1}L_{\frac{n_{12}-1}{2}}^{j+1+\mu_1+\mu_2}(\rho^2) &\text{ for }n_{12},j\text{ odd,}\\
 \displaystyle\sqrt{\frac{2\frac{n_{12}-1}{2}!}{\Gamma(\frac{n_{12}-1}{2}+j+\mu_1+\mu_2+2)}} e^{-\rho^2/2}\rho^{j+1}L_{\frac{n_{12}-1}{2}}^{j+1+\mu_1+\mu_2}(\rho^2) &\text{ for }n_{12}\text{ odd, }j\text{ even,}\\
 \displaystyle\sqrt{\frac{2\frac{n_{12}}{2}!}{\Gamma(\frac{n_{12}}{2}+j+\mu_1+\mu_2+1)}} e^{-\rho^2/2}\rho^{j}L_{\frac{n_{12}}{2}}^{j+\mu_1+\mu_2}(\rho^2) &\text{ for }n_{12}\text{ even, }j\text{ odd,}
\end{cases}
\end{align*}
and the angular wavefunctions, by
\begin{align*}
 F_j^{\mu_1,\mu_2}(\phi) &= \xi^+_j \Big[ P_{\frac{j}{2}}^{(\mu_2-1/2,\mu_1-1/2)}(\cos 2\phi) &\text{ for }n_{12},j\text{ even,}\\
	 & \qquad\qquad - \cos(\phi)\sin(\phi)P_{\frac{j}{2}-1}^{(\mu_2+1/2,\mu_1+1/2)}(\cos 2\phi)\Big] &\\
 F_j^{\mu_1,\mu_2}(\phi) &= \xi^+_j \Bigg[ \sqrt{\frac{\frac{j-1}{2}+1}{\frac{j-1}{2}+\mu_1+\mu_2+1}}P_{\frac{j+1}{2}}^{(\mu_2-1/2,\mu_1-1/2)}(\cos 2\phi) &\text{ for }n_{12},j\text{ odd,}\\
	 &  \qquad\qquad +\sqrt{\frac{\frac{j-1}{2}+\mu_1+\mu_2+1}{\frac{j-1}{2}+1}} \cos(\phi)\sin(\phi)P_{\frac{j-1}{2}}^{(\mu_2+1/2,\mu_1+1/2)}(\cos 2\phi)\Bigg] &\\
 F_j^{\mu_1,\mu_2}(\phi) &= \xi^-_j \Big[ \sin(\phi)P_{\frac{j}{2}}^{(\mu_2+1/2,\mu_1-1/2)}(\cos 2\phi) &\text{ for }n_{12}\text{  	odd, }j\text{ even,}\\
	 & \qquad\qquad + \cos(\phi)P_{\frac{j}{2}}^{(\mu_2-1/2,\mu_1+1/2)}(\cos 2\phi)\Big] &\\
 F_j^{\mu_1,\mu_2}(\phi) &= \xi^-_j \Bigg[ \sqrt{\frac{\frac{j-1}{2}+\mu_1+1/2}{\frac{j-1}{2}+\mu_2+1/2}}\sin(\phi)P_{\frac{j-1}			 	{2}}^{(\mu_2+1/2,\mu_1-1/2)}(\cos 2\phi) &\text{ for }n_{12}\text{ even, }j\text{ odd,}\\
	 & \qquad\qquad -\sqrt{\frac{\frac{j-1}{2}+\mu_2+1/2}{\frac{j-1}{2}+\mu_1+1/2}} \cos(\phi)P_{\frac{j-1}{2}}^{(\mu_2-1/2,\mu_1+1/2)}(\cos 2\phi)\Bigg] &
\end{align*}
where $P_n^{\alpha,\beta}(x)$ are the Jacobi polynomials and with
\begin{align*}
 \xi^+_j &=
\begin{cases}
 \displaystyle\sqrt{\frac{(\frac{j}{2})!\Gamma(\frac{j}{2}+\mu_1+\mu_2+1)}{2\Gamma(\frac{j}{2}+\mu_1+1/2)\Gamma(\frac{j}{2}+\mu_2+1/2)}}\quad &\text{for }j\text{ even,}\\
 \displaystyle\sqrt{\frac{(\frac{j+1}{2})!\Gamma(\frac{j+1}{2}+\mu_1+\mu_2+1)}{2\Gamma(\frac{j+1}{2}+\mu_1+1/2)\Gamma(\frac{j+1}{2}+\mu_2+1/2)}}\quad &\text{for }j\text{ odd,}
\end{cases}\\
\xi^-_j &=
\begin{cases}
  \displaystyle\sqrt{\frac{(\frac{j}{2})!\Gamma(\frac{j}{2}+\mu_1+\mu_2+1)}{2\Gamma(\frac{j}{2}+\mu_1+1/2)\Gamma(\frac{j}{2}+\mu_2+1/2)}}\quad &\text{for }j\text{ even,}\\
  \displaystyle\sqrt{\frac{(\frac{j-1}{2})!\Gamma(\frac{j-1}{2}+\mu_1+\mu_2+1)}{2\Gamma(\frac{j-1}{2}+\mu_1+1/2)\Gamma(\frac{j-1}{2}+\mu_2+1/2)}}\quad &\text{for }j\text{ odd.}
\end{cases}
\end{align*}

\subsection{Generating function} \label{GenFunWavefunctions}

We remind the reader that the coupled and uncoupled wavefunctions decompose into each other with the CGC as coefficients. Writing the $x$ and $y$ variables in polar coordinates, this decomposition is given by
\begin{align}\label{wavefunctionsDecomp}
 \Psi_{n_{12}}^{\mu_{12},\epsilon_{12}}(\rho,\phi) &= \displaystyle\sum\limits_{n1,n2}C^{n_1 n_2}_{n_{12}j} \psi_{n_1}^{\mu_1}(\rho \cos \phi)\psi_{n_2}^{\mu_2}(\rho \sin \phi).
\end{align}
To bring \eqref{wavefunctionsDecomp} in the form of a generating relation for the CGC, one needs to transform the polynomials on the right hand side into monomials of a single variable. This can be accomplished by considering the asymptotic expansion when $\rho \rightarrow \infty$ of \eqref{wavefunctionsDecomp}. We will first derive the asymptotic expansion of the left hand side and then proceed to treat the right hand side.

From \eqref{CoupledAsRadialAngular}, we have that the coupled wavefunctions can be written as the product of a radial and angular wavefunctions. The asymptotic expansion for $\rho \rightarrow \infty$ will thus only be applied to the radial wavefunctions. The leading term of a Laguerre polynomial is given by
\begin{align}\label{leadingLaguerre}
 L_n^\alpha(x) \longrightarrow \frac{(-1)^n}{n!}x^n.
\end{align}
Using the above, one obtains the following asymptotic expression for the radial wavefunctions
\begin{align}\label{RadialAsymptotic}
 P_{n_{12}}^{\mu_{12}}(\rho) \longrightarrow e^{-\rho^2/2} \Lambda_{n_{12}}^{\mu_{12}} \rho^{n_{12}+j},
\end{align}
where $\Lambda_{n_{12}}^{\mu_{12}}$ is given by
\begin{align*}
 \Lambda_{n}^{\mu} =
\begin{cases}
 \displaystyle\sqrt{\frac{2}{\Gamma(\frac{n+1}{2}+\mu)(\frac{n}{2})!}}(-1)^{n/2}\quad &\text{for } n \text{ even},\\
 \displaystyle\sqrt{\frac{2}{\Gamma(\frac{n}{2}+\mu+1)(\frac{n-1}{2})!}}(-1)^{\frac{n-1}{2}}\quad &\text{for } n \text{ odd}.
\end{cases}
\end{align*}

Consider now the right hand side of \eqref{wavefunctionsDecomp} which can be written explicitly, using \eqref{CartesianWavefunctions}, as
\begin{align}\label{WaveLefthandside}
 \displaystyle\sum\limits_{n1,n2}C^{n_1 n_2}_{n_{12}j} \psi_{n_1}^{\mu_1}(\rho \cos \phi)\psi_{n_2}^{\mu_2}(\rho \sin \phi) =  e^{-\rho^2/2} \displaystyle\sum\limits_{n=0}^{n_{12}+j}C^{n_1,n_2}_{n_{12}, j} H_{n}^{\mu_1}(\rho\cos\phi)H_{n_{12}+j-n}^{\mu_2}(\rho\sin\phi),
\end{align}
where \eqref{labelsEquation} has been used. We will not mind the exponential factor in the asymptotic expansion since it will be cancelled by the same factor on the left hand side of \eqref{wavefunctionsDecomp}. One then only needs to evaluate the leading term for the two generalized Hermite polynomials in the above sum. Knowing the leading term for Laguerre polynomials \eqref{leadingLaguerre} and using \eqref{generalizedHermite}, we obtain for the leading term
\begin{align}
  H_n^{\mu}(x) &\longrightarrow N_n^\mu x^n,
\end{align}
where the normalization $N_n^\mu$ is given by
\begin{align*}
 N_\mu(n) = 
\begin{cases}
  \left[\Gamma(\frac{n}{2}+\mu+1/2)(\frac{n}{2})!\right]^{-1/2} &n\text{ even,}\\
  \left[\Gamma(\frac{n+1}{2}+\mu+1/2)(\frac{n-1}{2})!\right]^{-1/2} &n\text{ odd.}
\end{cases}
\end{align*}
Using the above expression for the leading term of the generalized Hermite polynomials, one can obtain the asymptotic expansion of \eqref{WaveLefthandside}, omitting the exponential factor,
\begin{multline}\label{DecoupledAsymptotic}
 \displaystyle\sum\limits_{n=0}^{n_{12}+j}C^{n_1,n_2}_{n_{12}, j} H_{n}^{\mu_1}(\rho\cos\phi)H_{n_{12}+j-n}^{\mu_2}(\rho\sin\phi)\\
 \longrightarrow (\rho\sin\phi)^{n_{12}+j} \displaystyle\sum\limits_{n=0}^{n_{12}+j}C^{n_1, n_2}_{n_{12}, j} N_{\mu_1}(n)N_{\mu_2}(n_{12}+j-n)\cot^{n}\phi.
\end{multline}
Using \eqref{CoupledAsRadialAngular} and \eqref{RadialAsymptotic} on the left hand side of \eqref{wavefunctionsDecomp} while combining \eqref{WaveLefthandside} and \eqref{DecoupledAsymptotic} to express the right hand side, we are lead to the asymptotic expression for \eqref{wavefunctionsDecomp} when $\rho \rightarrow \infty$. Simplifying, this expression becomes
\begin{align*}
 \Lambda_{n_{12}}^{\mu_{12}} F_j^{\mu_1,\mu_2}(\phi)(\csc\phi)^{n_{12}+j} = \displaystyle\sum\limits_{n=0}^{n_{12}+j}C^{n_1, n_2}_{n_{12}, j} N_{\mu_1}(n)N_{\mu_2}(n_{12}+j-n)\cot^{n}\phi
\end{align*}
The right hand side is a power series of a single variable $z$ if we take $z=\cot\phi$. The left hand side, when expressed as a function of $z$ is thus a generating function for the $\mathfrak{osp}(1|2)$ algebra CGC. Let us connect the form of the generating function we just obtained to the form found in the preceeding section. We want to express
\begin{align*}
 F_j^{\mu_1,\mu_2}(\phi)(\csc\phi)^{n_{12}+j}
\end{align*}
as a hypergeometric series in $z=\cot\phi$. Note that
\begin{align*}
 (\csc\phi)^{n_{12}+j} = (1 + z^2)^{\frac{n_{12}+j}{2}}.
\end{align*}
The angular wavefunctiond need to be treated separately for each parity combination of $n_{12}$ and $j$.

\subsubsection{Case with $n_{12}$ and $j$ even}

The angular wavefunction in this case is given by
\begin{align*}
 F_j^{\mu_1,\mu_2}(\phi) = \xi^+_j \Big[ P_{\frac{j}{2}}^{(\mu_2-1/2,\mu_1-1/2)}(\cos 2\phi) - \cos(\phi)\sin(\phi)P_{\frac{j}{2}-1}^{(\mu_2+1/2,\mu_1+1/2)}(\cos 2\phi)\Big].
\end{align*}
One first exchanges the parameters of the Jacobi polynomials using
\begin{align*}
 P_n^{(\alpha,\beta)}(z) &= (-1)^n P_n^{(\beta,\alpha)}(-z)
\end{align*}
to obtain
\begin{align*}
 F_j^{\mu_1,\mu_2}(\phi) = \xi^+_j (-1)^{j/2}\Big[ P_{\frac{j}{2}}^{(\mu_1-1/2,\mu_2-1/2)}(-\cos 2\phi) + \cos(\phi)\sin(\phi)P_{\frac{j}{2}-1}^{(\mu_1+1/2,\mu_2+1/2)}(-\cos 2\phi)\Big].
\end{align*}
Using the definition of the Jacobi polynomials in terms of hypergeometric series, we can write the above as
\begin{align*}
 F_j^{\mu_1,\mu_2}(\phi) &=\xi^+_j (-1)^{\frac{j}{2}} \Bigg[ \frac{(\mu_1+1/2)_{j/2}}{(j/2)!}{}_2 F_1\left(\begin{matrix} -\frac{j}{2},\quad \mu_1+\mu_2+\frac{j}{2} \\ \mu_1+1/2 \end{matrix} ; \frac{1+\cos2\phi}{2}\right)\\
	   &+ \cos\phi\sin\phi\frac{(\mu_1+3/2)_{\frac{j}{2}-1}}{(\frac{j}{2}-1)!}{}_2 F_1\left(\begin{matrix} 1-\frac{j}{2},\quad \mu_1+\mu_2+\frac{j}{2}+1 \\ \mu_1+3/2 \end{matrix} ; \frac{1+\cos2\phi}{2}\right)\Bigg],
\end{align*}
which, with the help of the following identity for ${}_2 F_1$ hypergeometric series
\begin{align*}
 {}_2 F_1\left(\begin{matrix} a,\quad b \\ c \end{matrix} ; \frac{x}{x-1} \right) = (1-x)^{a} {}_2 F_1\left(\begin{matrix} a, \quad c-b \\ c \end{matrix} ; x \right),
\end{align*}
and with $z=\cot\phi$, can be expressed as
\begin{align*}
 F_j(z) &=\xi^+_j (-1)^{\frac{j}{2}}(1+z^2)^{-\frac{j}{2}}\Bigg[ \frac{(\mu_1+1/2)_{j/2}}{(j/2)!}{}_2 F_1\left(\begin{matrix} -\frac{j}{2},\quad -\mu_2-\frac{j}{2}+1/2 \\ \mu_1+1/2 \end{matrix} ; -z^2\right)\\
	   &+ z\frac{(\mu_1+3/2)_{\frac{j}{2}-1}}{(\frac{j}{2}-1)!}{}_2 F_1\left(\begin{matrix} 1-\frac{j}{2},\quad -\mu_2-\frac{j}{2}+1/2 \\ \mu_1+3/2 \end{matrix} ; -z^2\right)\Bigg].
\end{align*}
By factoring the common factors of both hypergeometric series, we recover the generating function
\begin{multline*}
 F_j^{\mu_1,\mu_2}(\phi)(\csc\phi)^{n_{12}+j} =\xi^+_j (-1)^{\frac{j}{2}}\frac{(\mu_1+1/2)_{j/2}}{(j/2)!}\Bigg[{}_2 F_1\left(\begin{matrix} -\frac{j}{2},\quad \frac{1}{2}-\frac{j}{2}-\mu_2 \\ \frac{1}{2}+\mu_1 \end{matrix} ; -z^2\right)\\
	   + \frac{jz}{2\mu_1+1}{}_2 F_1\left(\begin{matrix} 1-\frac{j}{2},\quad \frac{1}{2}-\frac{j}{2}-\mu_2 \\ \frac{3}{2}+\mu_1 \end{matrix} ; -z^2\right)\Bigg](1+z^2)^{\frac{n_{12}}{2}}.
\end{multline*}

\subsubsection{Other cases of parity of $n_{12}$ and $j$}

The calculations to connect the generating function to the known expressions in the other cases are very similar. We will thus only list the results with their respective normalizations for reference.

For $n_{12}$ even and $j$ odd, one obtains
\begin{multline*}
F_j^{\mu_1,\mu_2}(\phi)(\csc\phi)^{n_{12}+j} = \xi^-_j \frac{(-1)^{\frac{j-1}{2}}}{(\frac{j-1}{2})!}\sqrt{\frac{\frac{j-1}{2}+\mu_1+1/2}{\frac{j-1}{2}+\mu_2+1/2}} (\mu_1+1/2)_{\frac{j-1}{2}} \Bigg[{}_2 F_1\left(\begin{matrix} -\frac{j-1}{2},\quad -\frac{j}{2}-\mu_2 \\ \frac{1}{2}+\mu_1 \end{matrix} ; -z^2 \right)\\
	 -z\left(\frac{j+2\mu_2}{2\mu_1 + 1}\right) {}_2 F_1\left(\begin{matrix} -\frac{j-1}{2},\quad 1-\frac{j}{2}-\mu_2 \\ \frac{3}{2}+\mu_1 \end{matrix} ; -z^2 \right)\Bigg](1+z^2)^{\frac{n_{12}}{2}},
\end{multline*}
while for $n_{12}$ odd and $j$ even, we have
\begin{multline*}
F_j^{\mu_1,\mu_2}(\phi)(\csc\phi)^{n_{12}+j} = \xi^-_j \frac{(-1)^{\frac{j}{2}}}{(\frac{j}{2})!}(\mu_1+1/2)_{\frac{j}{2}}\Bigg[{}_2 F_1\left(\begin{matrix} -\frac{j}{2},\quad -\frac{j}{2}-\mu_2-\frac{1}{2} \\ \frac{1}{2}+\mu_1 \end{matrix} ; -z^2\right)\\
	 +z\left(\frac{j+2\mu_1+1}{2\mu_1+1}\right) {}_2 F_1\left(\begin{matrix} -\frac{j}{2},\quad \frac{1}{2}-\frac{j}{2}-\mu_2 \\ \frac{3}{2}+\mu_1 \end{matrix} ; -z^2\right)\Bigg](1+z^2)^{\frac{n_{12}-1}{2}}
\end{multline*}
and finally, for $n_{12}$ and $j$ odd, one has
\begin{multline*}
F_j^{\mu_1,\mu_2}(\phi)(\csc\phi)^{n_{12}+j} = \xi^+_j \frac{(-1)^{\frac{j+1}{2}}}{(\frac{j+1}{2})!}\sqrt{\frac{\frac{j-1}{2}+1}{\frac{j-1}{2}+\mu_1+\mu_2+1}}(\mu_1+1/2)_{\frac{j+1}{2}}\Bigg[{}_2 F_1\left(\begin{matrix} -\frac{j+1}{2},\quad -\frac{j}{2}-\mu_2 \\ \frac{1}{2}+\mu_1 \end{matrix} ; -z^2\right)\\
	  - z \left(\frac{j+2\mu_1+2\mu_2+1}{2\mu_1+1}\right) {}_2 F_1\left(\begin{matrix} -\frac{j-1}{2},\quad -\frac{j}{2}-\mu_2 \\ \frac{3}{2}+\mu_1 \end{matrix} ; -z^2\right)\Bigg](1+z^2)^{\frac{n_{12}-1}{2}},
\end{multline*}
where in all cases, we have, as shown in section \ref{GenFunWavefunctions}, that
\begin{align*}
 \Lambda_{n_{12}}^{\mu_{12}} F_j^{\mu_1,\mu_2}(\phi(z))(\csc\phi(z))^{n_{12}+j} = \displaystyle\sum\limits_{n=0}^{n_{12}+j}C^{n_1, n_2}_{n_{12}, j} N_{\mu_1}(n)N_{\mu_2}(n_{12}+j-n)z^{n},
\end{align*}
with $\phi(z) = \cot^{-1} z$.

\section{Conclusion}
We have shown that the approach first introduced in \cite{3nj} to calculate Clebsch-Gordan coefficients admits a generalization to algebras with a twisted coproduct. It required the relation \eqref{xyzEquation} between the eigenvalues of the coupled and uncoupled coherent states which is quadratic because the twisted element in the coproduct involves an involution. This suggest that our extension might be applicable to a variety of algebras with that property. The connection between $\mathfrak{osp}(1|2)$ and the dynamical algebra of a parabosonic oscillator has also been exploited to derive the generating function by analytical manipulations of the wavefunction realizations of this algebra. The results of this paper suggest it may be possible to derive generating functions for higher order coupling coefficients, such as the Racah coefficients. The Racah coefficients of $\mathfrak{osp}(1|2)$ are known to be the Bannai-Ito polynomials, for which a generating function has not yet been derived. With the existence of several multivariable generalizations of orthogonal polynomials, it would also be of interest to explore the applicability of the approach to the multivariate cases.

\newpage
\section*{Acknowledgments}
The authors would like to thank Vincent X. Genest for stimulating discussions. The research of Geoffroy Bergeron was supported through scholarships of the Natural Science and Engineering Research of Canada (NSERC) and of the Fond de Recherche du Québec - Nature et Technologies (FRQNT). The research of Luc Vinet is supported in part by the Natural Science and Engineering Research Council of Canada (NSERC).

\bibliographystyle{plain}

\end{document}